\documentclass[twocolumn]{aastex631}

\newcommand{\R}{\textcolor{red}}
\newcommand{\Bf}{\textbf}
\newcommand{\und}{\underline}
\usepackage{csquotes}
\usepackage{dirtytalk}
\usepackage{multirow}
\usepackage{subfigure}

\usepackage{amsmath}

\shorttitle{Survey of Ten Exoplanetary Atmospheres}
\shortauthors{}

\graphicspath{{./}{figures/}}

\begin{document}

\title{A Homogeneous Survey of JWST MIRI Transmission Spectra of 10 Exoplanets}

\author[0000-0002-0931-735X]{Martin Binet}
\affiliation{Institute of Astronomy, University of Cambridge, Madingley Road, Cambridge CB3 0HA, UK}

\author[0000-0002-4869-000X]{Nikku Madhusudhan}
\affiliation{Institute of Astronomy, University of Cambridge, Madingley Road, Cambridge CB3 0HA, UK}

\author[0000-0002-0931-735X]{Måns Holmberg}
\affiliation{Space Telescope Science Institute, 3700 San Martin Drive, Baltimore, MD 21218, USA}

\author[0000-0002-2705-5402]{Subhajit Sarkar} 
\affiliation{School of Physics and Astronomy, Cardiff University, The Parade, Cardiff CF24 3AA, UK}

\correspondingauthor{Martin Binet, Nikku Madhusudhan}
\email{mb2672@cam.ac.uk, nmadhu@ast.cam.ac.uk} 

\begin{abstract}
The launch of JWST has enabled unprecedented atmospheric characterization of exoplanets. In particular, the JWST-MIRI instrument in the mid-infrared enables probing of molecular species that are harder to detect in the near-infrared. This has come to the fore with the first MIRI-LRS observation of a candidate hycean world, K2-18~b, with potential signs of dimethyl sulfide in its atmosphere. The statistical significance of this result has been debated in subsequent works, arguing instead for the possibility of random noise, instrumental systematics, or a different molecule. The same applies to similar spectra reported for several other exoplanets using MIRI-LRS and interpreted with similar retrieval architectures. In this work, we distinguish between these possibilities with a homogeneous survey of MIRI-LRS transmission spectra of 10 diverse exoplanets, from temperate sub-Neptunes to hot rocky planets and gas giants. We first perform a comparative assessment of the spectra, considering fits with featureless models and pairwise correlations, followed by extensive atmospheric retrievals. With the retrievals, we search for over 150 trace species in the atmosphere of each planet. We find hints of complex molecules in the temperate sub-Neptunes and candidate hycean worlds K2-18~b, TOI-732~c and TOI-270~d, similar to previous works, but not in the other seven planets. Our results indicate that the observed spectral features are likely due to molecular absorption rather than instrument systematics or noise, and underscore these temperate sub-Neptunes as a unique class of planets. More observations are required to robustly identify the specific molecules and constrain the underlying chemical processes, biotic or abiotic. 
\end{abstract}

\section{Introduction} 
\label{sec:intro}

The launch of the James Webb Space Telescope (JWST; \citealt{Gardner_Mather_2023}) has enabled atmospheric characterization of exoplanets with greater precision than ever before. With transit spectroscopy, some of the key discoveries have been robust detections of molecular species in near-infrared observations of a  wide range of planets, from hot gas giants to temperate sub-Neptunes. For gas giants, these include H$_2$O, CO$_2$, CO and SO$_2$ in WASP-39~b \citep{Ahrer_Stevenson_2023, Alderson2023, Feinstein_Radica_2023}, H$_2$S in HD~189733~b \citep{Fu_Welbanks_2024}, CH$_4$ in WASP-80~b \citep{Bell_Welbanks_2023}, and potentially OCS in the young planet V1298~Tau~b \citep{Barat_Desert_2025}. Several sub-Neptunes (1.8$R_\oplus \lesssim R_p \lesssim 4R_\oplus$) have also shown molecular features in H$_2$/He-dominated atmospheres, including CH$_4$ and CO$_2$ in K2-18~b \citep{Madhusudhan_Sarkar_2023} and TOI-270~d \citep{Benneke2024, Holmberg2024}, as well as H$_2$O in TOI-421~b \citep{Davenport2025}. Even hints of secondary atmospheres have been inferred in some planets, including LHS~1140~b \citep{Damiano_Bello-Arufe_2024}, GJ~1214~b \citep{Kempton2023}, GJ~9827~d \citep{Piaulet-Ghorayeb_Benneke_2024} and L~98-59~d \citep{Banerjee_Barstow_2024, Gressier2024}, the latter being subsequently supported by ground-based observations \citep{Cheverall_Madhusudhan_2026}. More detailed reviews of JWST advances in exoplanet atmosphere characterization can be found in \cite{Espinoza_Perrin_2026} and \cite{Madhusudhan_Holmberg_2025}, the latter being focused on sub-Neptunes.

Although NIR transit observations benefit from a higher signal-to-noise ratio (SNR) due to stronger stellar flux, those with the JWST Mid Infrared Instrument (MIRI) can give precious complementary information. 
In particular, the MIRI Low Resolution Spectrograph (MIRI-LRS, \citealt{Kendrew2015}) instrument provides a broad wavelength range (5-12 $\mu$m), which encompasses spectral features of various molecules and condensate species \citep{Wakeford2015, Pinhas2017, madhusudhan2026}. This was demonstrated for several giant exoplanets starting with the JWST Transit Early Release Science (ERS) program for WASP-39~b. In this case, a transmission spectrum obtained with MIRI-LRS led to an inference of SO$_2$ in WASP-39~b, complementing previous findings with the NIR spectra \citep{Powell2024}. Similar inferences were also made for other hot Jupiters, including H$_2$O and SO$_2$ in WASP-107~b and silicate clouds in WASP-17~b \citep{Grant_Lewis_2023}.  

Most recently, the importance of MIRI-LRS has been demonstrated for transmission spectroscopy of temperate exoplanets, with the first mid-infrared spectrum of a habitable-zone sub-Neptune, K2-18~b \citep{Madhusudhan_Constantinou_2025}, followed by other temperate sub-Neptunes \citep{Rigby_Madhusudhan_2025a, Holmberg_Madhusudhan_2026}. The MIRI-LRS spectrum of K2-18~b was used to infer tentative evidence for dimethyl sulfide (DMS) and/or dimethyl disulfide (DMDS), both potential biomarkers, at $\sim$$3\sigma$ (ln B $\sim$ 3). The reported evidence is stronger than the previous hint of DMS reported using NIR observations with JWST NIRISS and NIRSpec G395H, which was affected by the degeneracy with CH$_4$ and CO$_2$ and strong dependence on detector offsets \citep{Madhusudhan_Sarkar_2023}. The broader wavelength coverage and single detector, combined with the multiple strong features of DMS in the mid-infrared, rendered it more observable with MIRI-LRS. 

However, as alluded to in \citet{Madhusudhan_Constantinou_2025}, the low SNR of this single MIRI-LRS observation allowed for other interpretations in the form of unknown systematics \citep{Stevenson2025} or other chemical species besides DMS and DMDS \citep{Jaziri_Drant_2025, Luque_Piaulet-Ghorayeb_2025, Pica-Ciamarra_Madhusudhan_2026, Welbanks_Nixon_2026}. If the observed feature is indeed from the planet, then a comprehensive exploration of over 650 molecules suggests that DMS is the most likely explanation of the data \citep{Pica-Ciamarra_Madhusudhan_2026}. Similar interpretations of complex molecules are possible for other temperate sub-Neptunes such as TOI-732~c \citep{Rigby_Madhusudhan_2025a} and TOI-270~d \citep{Holmberg_Madhusudhan_2026}. However, it is important to consider the possibility of unknown systematics which, in turn, could affect other observations \citep{Stevenson2025}. 

The consideration of unknown systematics is particularly important given the diverse range of exoplanets being observed with JWST MIRI. Indeed, the attributes of low SNR and detection significance observed for K2-18~b extend beyond temperate sub-Neptunes to similar observations of larger and hotter exoplanets. For example, as noted above, one of the earliest exoplanet observations with MIRI-LRS was of a transmission spectrum of the hot jupiter WASP-17~b with an inference of silicate clouds in the atmosphere \citep{Grant_Lewis_2023}. The reported model preference for silicate clouds over a generic gray-cloud setup was $2.6\sigma$ (ln B $\sim$ 2) which is comparable to the inference of DMS/DMDS in K2-18~b \citep{Madhusudhan_Constantinou_2025}, with comparable retrieval architectures considering three to four relevant gas-phase molecules to compute the model preferences. It is to be noted, however, that the inference for WASP-17~b was obtained with one of the two data reductions considered, so the robustness to different pipeline choices remains open. Similarly, another relevant example is the case of WASP-39~b observed in the JWST ERS program for which a MIRI-LRS spectrum was used to infer the presence of SO$_2$ at $\gtrsim$ 3$\sigma$ (ln B $\gtrsim$ 3), and with a retrieval architecture including only two species H$_2$O and SO$_2$ \citep{Powell2024} . Therefore, any competing hypotheses invoked to explain the spectrum of K2-18~b, such as unknown systematics or unexpected molecules, would also apply to these observations of hot Jupiters. 

It is, therefore, imperative to address these various interpretations with a homogeneous survey of JWST MIRI transmission spectra for a diverse range of targets. Such a survey is required to effectively disentangle the  instrumental systematics and random noise from potential astrophysical signals. For example, if there are strong instrumental systematics, we expect to find similar signals across the datasets independent of the targets. Similarly, if random noise dominates the spectra, we do not expect to see significant correlations between different observations. On the other hand, if neither instrumental nor random noise dominate the spectra, the diverse datasets open an exciting avenue for comparative characterization of the diverse atmospheres in the mid-infrared. 

To achieve this goal, in this work we conduct a comprehensive and homogeneous survey of 10 diverse exoplanets observed with JWST-MIRI using transmission spectroscopy, encompassing all the datasets published to date, spanning gas giants, sub-Neptunes and rocky exoplanets. In what follows, we discuss the target sample and observations along with a correlation analysis in section~\ref{sec:observations}. We then present our atmospheric retrieval setup in section~\ref{sec:retrieval_setup}, and the results in section~\ref{sec:results}. We summarize and discuss our findings in section~\ref{sec:discussion}.

\section{Target Sample and Observations} 
\label{sec:observations}
We start by listing the wide range of planets that constitute our target sample, providing information on their bulk characteristics and atmospheric properties reported previously using transit spectroscopy. We then perform a homogeneous comparison of all the spectra, by fitting them with a flat model, followed by pairwise comparisons using the Pearson correlation coefficient. 

\subsection{Planet Sample}
\label{subsec:planets}
\begin{figure}
\centering
	\includegraphics[width=0.47\textwidth]{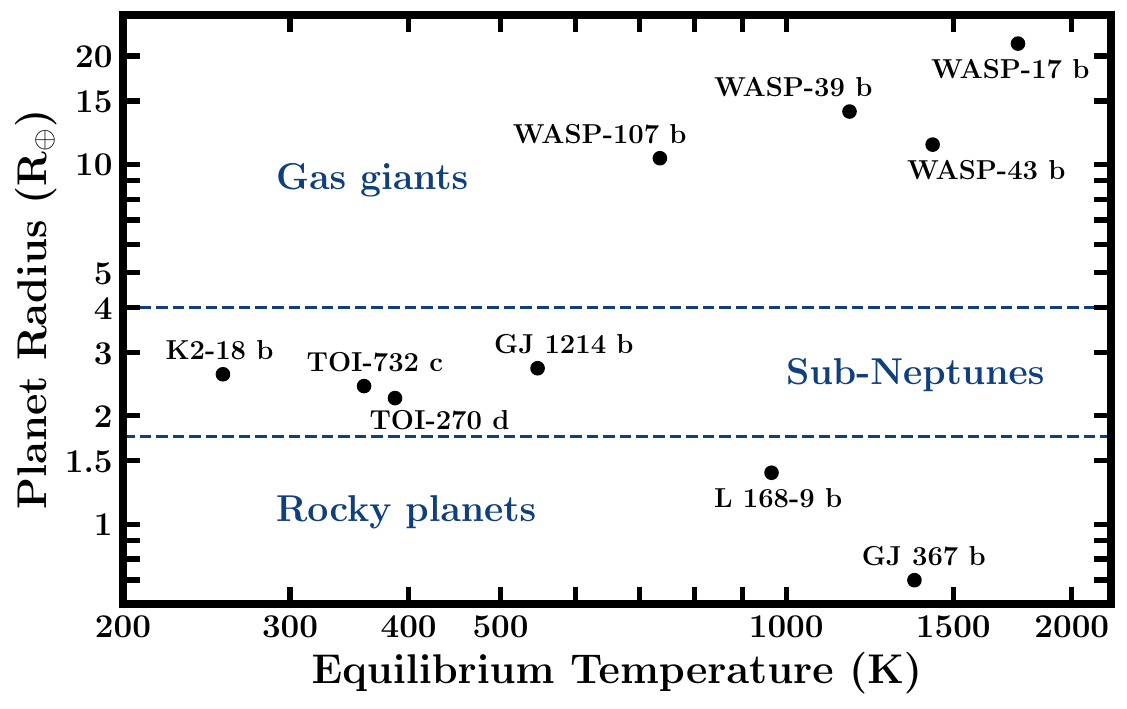}
     \caption{Sizes and temperatures of the ten planets studied in this work. The sample includes a wide variety, with four gas giants, four sub-Neptunes and two rocky planets, spanning temperate to hot conditions.}
     \label{fig:planets-plot} 
\end{figure}

In this work, we focus on transmission spectra observed with JWST MIRI-LRS. There are 11 planets that have such publicly available spectra published in the literature, listed in Table~\ref{tab:planets-list} and Figure~\ref{fig:planets-plot}: K2-18~b, TOI-732~c, TOI-270~d, GJ~1214~b, WASP-107~b, L~168-9~b, WASP-39~b, GJ~367~b, WASP-43~b, WASP-17~b and K2-22~b. 
From these, we discard the disintegrating rocky planet K2-22~b \citep{Sanchis-Ojeda_Rappaport_2015} that has a highly variable transit depth in its MIRI spectrum \citep{Tusay_Wright_2025}. This leaves us with 10 planets, 7 of which are too large and/or hot ($T_{eq} \geq 500 K$) to be habitable. These will be compared with the temperate ($T_{eq} \leq 400K$) sub-Neptunes K2-18~b, TOI-270~d and TOI-732~c which are potentially habitable as candidate hycean worlds \citep{Madhusudhan_Piette_2021}. We first briefly summarize the previously published results from transit observations of these planets. 

\subsubsection{Uninhabitable Planets}
\label{subsubsec:uninhabitable-planets}
We start by introducing the uninhabitable planets in the sample. First, we have the hot super-Earth L~168-9~b \citep{Astudillo-Defru_Cloutier_2020}. Its MIRI transmission spectrum was first reduced in \cite{Bouwman2023}. Then, \cite{Alam2025} independently reduced the MIRI observation, and also observed a transit in the NIR with JWST NIRSpec G395H ($2.9-5.1\mu m$). This ruled out any clear low-metallicity atmosphere, but the presence of a cloudy or metal-dominated atmosphere remains a possibility. 

There is another rocky planet in this list, GJ~367~b, discovered in 2021 \citep{Lam_Csizmadia_2021}. Its only atmospheric observation was the MIRI phase curve presented in \cite{Zhang_Hu_2024}. As expected from its small size and high equilibrium temperature (0.68 $R_\oplus$, 1360 K), this data was consistent with the planet having a zero-albedo and no heat redistribution, hence being a bare rock. This includes the transit spectrum being consistent with a flat line (see section~\ref{subsubsec:flat-line-fits}), but we still include it in our study for completeness.

Then, GJ~1214~b \citep{charbonneau2009} is a warm Sub-Neptune that was first observed with JWST in \cite{Kempton2023}. This observation was a MIRI phase curve, that found the planet to likely have a high-metallicity atmosphere with clouds or hazes, potentially water-dominated. This was also found in \cite{Gao_Piette_2023}, by combining the transit from this MIRI phase curve with Spitzer and Hubble Space Telescope (HST) transits, from \cite{Fraine_Deming_2013} and \cite{Kreidberg_Bean_2014_GJ1214b} respectively. Later, it was also observed with NIRSpec, and new studies still found a high-metallicity atmosphere that also showed potential CO$_2$ and CH$_4$ features \citep{Schlawin_Ohno_2024, Ohno_Schlawin_2025}. CO$_2$ was then also inferred in a CRIRES$^+$ high-resolution spectrum \citep{Nortmann_Cont_2026}.
As a side note, \cite{Orell-Miquel_Murgas_2022} also found tentative evidence for helium escape with the ground-based spectrograph CARMENES.

\begin{table*}
\centering
\renewcommand{\arraystretch}{1.05} 
\setlength{\tabcolsep}{2pt}
\begin{tabular}{lccccl}
\hline \hline
Planet             & R$_p$/R$_\oplus$ & M$_p$/M$_\oplus$ &T$_{eq}$/K &R$_\star$/R$_\odot$ & Previous findings \\\hline
K2-18~b              & 2.56 & $8.63 \pm 1.35^{[1]}$  & 280   & $0.44$   & \Bf{DMS$^a$}, CH$_4$, CO$_2$   \\
TOI-732~c            & 2.41 & $8.04 \pm 0.50^{[2]}$  & 360   & 0.38   & \Bf{Isobutene$^a$, 1-pentene$^a$}, CH$_4$ \\ 
TOI-270~d            & 2.19 & $4.78 \pm 0.43^{[3]}$  & 390   & 0.38   & CH$_4$, CO$_2$, H$_2$O$^a$, CS$_2$$^a$  \\ 
GJ~1214~b$^{PC/+NIR}$& 2.73 &$8.41 \pm 0.36^{[4]}$  & 570   & 0.22   & \Bf{H$_2$O$^a$}, CO$_2^a$, CH$_4^a$, He$^a$ (clouds and/or high Z)  \\
WASP-107~b$^{+HST}$& 9.70 & $30.5 \pm 1.59^{[5]}$  & 750   & 0.66   & \Bf{H$_2$O, SO$_2$, Si clouds, H$_2$S, NH$_3$, CO}, CO$_2$, He, CH$_4$\\
L~168-9~b$^{+NIR}$ & 1.39 & $4.07 \pm 0.45^{[6]}$  & 960   & 0.60   & None (likely bare rock or high Z)         \\ 
WASP-39~b          & 14.3 & $89.3 \pm 9.54^{[7]}$  & 1150  & 0.93   & \Bf{H$_2$O, SO$_2$}, CO$_2$, CO, K, Na, H$_2$S$^a$, SiO$^a$ \\
GJ~367~b$^{PC}$    & 0.68 & $0.63 \pm 0.05^{[8]}$  & 1360  & 0.46   & None (likely bare rock)  \\
WASP-43~b$^{PC}$   & 11.3 & $652 \pm 16.5^{[9]}$   & 1400  & 0.66   & \Bf{H$_2$O, NH$_3$, CO$^a$}           \\ 
WASP-17~b$^{+HST}$ & 20.8 & $154 \pm 10.2^{[10]}$   & 1750  & 1.58   & \Bf{H$_2$O, Si clouds$^a$}, CO$_2$, H$^-$, K, Na  \\
K2-22~b$^{PC}$   &$\leq$0.71&$\leq445^{[11]}$&2100   & 0.57  & Disintegrating planet (discarded) \\ \hline
\end{tabular}
\caption{Planets with published MIRI transmission spectra. The equilibrium temperatures T$_{eq}$ are given for a Bond albedo of zero, and full heat redistribution.
We list the species previously inferred in each atmosphere, starting with those found in the MIRI data \Bf{in bold}.  
$^{PC}$ The transit was studied as part of a phase curve observation. 
$^{+NIR}$ The transit was studied jointly with JWST near-infrared data; $^{+HST}$ with Hubble data. $^a$~Inferences below the threshold for strong evidence (see Appendix~\ref{app:sigma-conversion}).  Wherever multiple degenerate species are reported, we only include those inferred in both MIRI and NIR data. \newline
References for previous findings are in section~\ref{subsec:planets}. Mass references: [1] \cite{Cloutier_Astudillo-Defru_2019}; [2] \cite{Bonfanti2024}; [3] \cite{Van-Eylen_Astudillo-Defru_2021}; [4] \cite{Mahajan_Eastman_2024}; [5] \cite{Wu_Zhang_2026}; [6] \cite{Hobson_Bouchy_2024}; [7] \cite{Faedi_Barros_2011}; [8] \cite{Goffo_Gandolfi_2023}; [9] \cite{Bonomo_Desidera_2017}; [10] \cite{Anderson_Smith_2011}; [11] \cite{Sanchis-Ojeda_Rappaport_2015}.
}
\label{tab:planets-list}
\end{table*}

Next, the hot Saturn WASP-107~b \citep{anderson2017} was first observed in transit with Hubble's WFC3 instrument ($0.9-1.6 \mu m$), and found to have a hydrogen-dominated atmosphere (as expected for a giant planet, and this is also true for the following planets) with water features \citep{Kreidberg_Line_2018} and escaping helium \citep{Spake_Sing_2018}. Then, \cite{Dyrek_Min_2024} presented a MIRI observation and performed joint retrievals with the Hubble data. They confirmed H$_2$O, and inferred both SO$_2$ and silicate clouds, all at $\geq 5\sigma$, as well as $2-5 \sigma$ evidence for H$_2$S, NH$_3$ and CO. Finally, H$_2$O, SO$_2$, NH$_3$ and CO were further confirmed by NIRCam \citep{Welbanks_Bell_2024} and NIRSpec G395H \citep{Sing_Rustamkulov_2024}, which also found CO$_2$ at $\geq20\sigma$ and CH$_4$ at $\approx 5\sigma$, although these works did not infer silicate clouds nor H$_2$S. Finally, \cite{Krishnamurthy_Carteret_2025} presented a NIRISS observation which confirmed the escaping helium detection. 

Next in our list, the hot Jupiter WASP-17~b \citep{Anderson_Hellier_2010} was found to have water features with Hubble WFC3 \citep{Mandell_Haynes_2013} and STIS ($0.3-1.0 \mu m$) \citep{sing2016}, the latter also finding K and Na features. Then, \cite{Alderson_Wakeford_2022} reanalyzed these datasets and also inferred $\approx3-4\sigma$ evidence for CO$_2$. Finally, the water absorption was confirmed with JWST MIRI combined with HST \citep{Grant_Lewis_2023} and NIRISS \citep{Louie_Mullens_2025}, along with tentative evidence for silicate clouds in the former, and H$^-$ in the latter. WASP-17~b also has an eclipse MIRI observation \citep{Valentine_Wakeford_2024}, which marginally further confirmed the water features.

As for WASP-39~b \citep{Faedi_Barros_2011}, it is one of the planets most studied by JWST in transit, with multiple near-infrared instruments: NIRISS \citep{Feinstein_Radica_2023}, NIRSpec G395H \citep{Alderson2023, Esparza-Borges_Lopez-Morales_2023, Grant_Lothringer_2023, Niraula_De-Wit_2023}, NIRSpec PRISM \citep{Constantinou2023, JWST_ERS2023, Rustamkulov2023}, NIRCam \citep{Ahrer_Stevenson_2023}, or even combining multiple spectra \citep{Carter2024, Lueber_Novais_2024, Sarkar2024, Ma_Saba_2026}. Together, these studies found spectral contributions from H$_2$O, SO$_2$, CO, CO$_2$, K and Na, as well as tentative evidence for H$_2$S and SiO. H$_2$O and SO$_2$ were also inferred with the MIRI observation of \cite{Powell2024}.

Finally, WASP-43~b \citep{Hellier_Anderson_2011} was first found to have water vapor features with HST WFC3 \citep{Kreidberg_Bean_2014_WASP-43b}. This was then confirmed with a MIRI phase curve observation in \cite{Bell_Crouzet_2024}, which was then reanalyzed in \cite{Yang_Hammond_2024} to also infer NH$_3$ and CO at $\approx3-4\sigma$.

We note that from these seven planets, WASP-39~b is the only one for which the MIRI transit data have been previously studied alone. This means that our retrieval results in section~\ref{subsec:results-uninhabitable} are directly comparable with \cite{Powell2024}.

\subsubsection{Habitable-zone Sub-Neptunes}
Now, we present the three temperate sub-Neptunes of the sample, which were predicted to be potential hycean worlds. First, K2-18~b \citep{Montet2015} was initially observed with Hubble WFC3 \citep{Benneke2019, Tsiaras_Waldmann_2019}. 
The spectrum was first inferred to show a hydrogen-dominated atmosphere and water vapour absorption. The observation was re-analyzed in \citep{Madhusudhan_Nixon_2020} which arrived at the same conclusion. Other studies noted degeneracies with CH$_4$ \citep{Blain_Charnay_2021, Bezard_Charnay_2022} and starspots \citep{Barclay_Kostov_2021}.
However, the planet was later observed with JWST NIRISS and NIRSpec G395H \citep{Madhusudhan_Sarkar_2023}, and the apparent water features turned out in fact to be CH$_4$, inferred at $\approx 5 \sigma$. CO$_2$ was also found at $\approx 3 \sigma$, which was contested in \cite{Schmidt_MacDonald_2025}, but later supported at $\approx 4 \sigma$ by additional data by \cite{hu_water-rich_2025}. This presence of both CH$_4$ and CO$_2$, but apparent depletion of CO, NH$_3$ and H$_2$O pointed toward the possibility of K2-18~b being hycean: an ocean world with a thin ($\leq 100$ bar) atmosphere on top \citep{Madhusudhan_Piette_2021}. Potential traces of DMS were also inferred, both from this NIR observation, but also from the later MIRI transit \citep{Madhusudhan_Constantinou_2025}, although there it was found to be degenerate with dimethyl disulfide (DMDS). As discussed in section~\ref{sec:intro} and \cite{Pica-Ciamarra_Madhusudhan_2026}, some other works suggested alternate explanations  \citep{Luque_Piaulet-Ghorayeb_2025, Stevenson2025, Taylor2025, Welbanks_Nixon_2026}, but the possibility of DMS was still supported considering all available data \citep{Pica-Ciamarra_Madhusudhan_2026, hu_water-rich_2025}.

Next, TOI-270~d \citep{gunther2019} was again first observed with HST WFC3 \citep{Mikal-Evans2023}, also finding a hydrogen-dominated atmosphere with water vapour. Then, with JWST NIRSpec G395H and NIRISS data \citep{Benneke2024, Holmberg2024, Felix_Kitzmann_2025, Constantinou2026}, the main species found were again CH$_4$ and CO$_2$, although this time H$_2$O was still favored at $\approx2\sigma$, along with CS$_2$ and potentially other species that are very degenerate (including C$_2$H$_6$, NH$_3$, CH$_3$F, CH$_3$Cl, H$_2$CS, CS, and DMS). 
These results have been interpreted as TOI-270~d being either a hycean planet \citep{Holmberg2024, Rigby_Madhusudhan_2025b, Constantinou2026}, a mixed-envelope planet \citep{Benneke2024}, or a magma ocean planet \citep{Nixon_Somers_2025}. Then, \cite{Holmberg_Madhusudhan_2026} analyzed the MIRI-LRS transmission spectrum of this planet and found evidence for excess absorption in the mid-infrared that could be explained by multiple species with degenerate spectral features, albeit different from those inferred in the NIR. 

Finally, TOI-732~c \citep{cloutier2020} was, on the contrary, never observed with HST. Its first atmospheric characterization was with ground-based spectroscopy, specifically with the IGRINS instrument of Gemini-South \citep{Cabot_Madhusudhan_2024}. This study found marginal evidence ($\approx2\sigma$) for CH$_4$. Then, \cite{Rigby_Madhusudhan_2025a} presented three JWST transits, one from each of NIRISS, NIRSpec G395H, and MIRI. This work confirmed the presence of CH$_4$ in a hydrogen-dominated atmosphere, but did not find substantial evidence for other major molecules. However, it found potential evidence for more complex molecules, including isobutene and 1-pentene which were independently favored in both the NIR and MIRI data.

\subsection{Datasets and Metrics}
For each planet, there has been only one MIRI transit observation. However, apart from GJ 1214~b and GJ~367~b, each observation has been reduced with at least two different pipelines, among \texttt{Eureka!} \citep{Bell2022}, \texttt{Tiberius} \citep{Kirk_Wheatley_2017, Rustamkulov2023}, \texttt{CASCADE} \citep{Carone_Molliere_2021, Bouwman2023}, \texttt{ExoTiC-MIRI} \citep{grant_david_2023}, \texttt{TEATRO} \citep{Dyrek_Min_2024, Crouzet_Edwards_2025}, \texttt{JExoRES} \citep{holmberg2023}, \texttt{JexoPipe} \citep{Sarkar2024}, \texttt{exoTEDRF} \citep{Radica_2024} and \texttt{SPARTA} \citep{Kempton2023}. To ensure that our work is robust to the pipeline choices, we use spectra from at least two different pipelines per planet when available and list them in Table~\ref{tab:spectra-list} and Table~\ref{tab:extra-spectra}. Because WASP-107~b has two spectra obtained with the \texttt{Eureka!} pipeline, we will refer to the one from \cite{Welbanks_Bell_2024} as the W24 spectrum throughout this work.

\subsubsection{Wavelength Ranges}
\label{subsubsec:wavelength-ranges}
We consider the wavelength ranges of the spectra based on previous works as follows. 
First, for the \texttt{JExoRES} and \texttt{JexoPipe} spectra of K2-18~b, TOI-270~d and TOI-732~c, we use the same wavelength ranges as the retrievals of \cite{Pica-Ciamarra_Madhusudhan_2026}, \cite{Rigby_Madhusudhan_2025a} and \cite{Holmberg_Madhusudhan_2026} respectively. 
Then, the three K2-18~b spectra from \cite{Luque_Piaulet-Ghorayeb_2025}, obtained with the reduction pipelines \texttt{SPARTA}, \texttt{exoTEDRF} and \texttt{Eureka!}, were plotted in their Figure 1. On that figure, we see that the blue-most data point at 5.125 $\mu m$ is somewhat inconsistent between the three spectra: $2858_{-40}^{+40}$, $3006_{-46}^{+46}$ and $3242_{-97}^{+97}$ ppm respectively. This corresponds to a discrepancy of 3.7$\sigma$ between \texttt{SPARTA} and \texttt{Eureka!}. All other data points are within $\lesssim 2 \sigma$ between the three reductions. Thus, we decide to remove the 5.125 $\mu m$ data point.
For the other spectra, we remove any data outside of the 5-12 $\mu$m range, in order to be consistent with the temperate sub-Neptunes: the first data point of the L~168-9~b spectrum, and the last 12 data points of the WASP-107~b spectrum. Finally, we note that all the spectral resolutions are kept as provided by each reference.
We summarize the wavelength ranges in Table~\ref{tab:spectra-list}. 

\subsubsection{Flat Line Fit}
\label{subsubsec:flat-line-fits}

\begin{table*}
\renewcommand{\arraystretch}{1.05} 
\setlength{\tabcolsep}{5.5pt}
\centering
\begin{tabular}{llccccccccccc}
\hline \hline
Planet & Pipeline & N$_p$  & $\lambda$ range & $\chi^2_{flat}$ & $\chi_{r,flat}^2$ & p$_{flat}$ & $\chi^2_{base}$ & n$_c$ & $\chi^2_{r,base}$ & p$_{base}$ & ln Z$_{base}$ & Ref \\ \hline
GJ~ 1214~b& \texttt{SPARTA}  & 14        & 5.28-11.75 & 25.5 & 1.96 & 0.02             & 23.8 & 3 & 2.16 & 0.01             & 103.59 & [1] \\ \hline
GJ~367~b  & \texttt{SPARTA}  & 11        & 5.36-11.45 & 17.3 & 1.73 & 0.07             & 16.8 & 2 & 1.87 & 0.05             & 94.88  & [2]\\ \hline
L~168-9~b & \texttt{Eureka!} & 47$^a$ & 5.01-11.86 & 75.3 & 1.64 & 4$\cdot 10^{-3}$ & 74.1 & 2 & 1.65 &4$\cdot$10$^{-3}$ & 366.43 & [3] \\  
           & \texttt{CASCADE$^{\times0.8}$}& 47$^a$ & 5.01-11.86 & 62.5 & 1.36 &0.05 & 52.8 & 2 & 1.17 & 0.20             & 384.18 & [3] \\ 
           & \texttt{ExoTiC}  & 47$^a$ & 5.01-11.86 & 52.8 & 1.15 & 0.23             & 49.3 & 3 & 1.12 & 0.27             & 379.79 & [4] \\ \hline
WASP-107~b & \texttt{Eureka!} & 46$^b$& 5.06-11.83 & 613  & 13.6 & $\leq$$10^{-6}$  & 74.8 & 9 & 2.02 & 2$\cdot$10$^{-4}$& 310.69 & [5] \\ \hline
WASP-17~b  & \texttt{Eureka!} & 28        & 5.13-11.88 & 55.2 & 2.04 & 1$\cdot 10^{-3}$ & 29.3 & 3 & 1.17 & 0.25             & 186.41 & [6] \\
           & \texttt{ExoTiC}  & 28        & 5.13-11.88 & 39.1 & 1.45 & 0.06             & 23.2 & 3 & 0.93 & 0.57             & 189.21 & [6] \\ \hline
WASP-39~b  & \texttt{Eureka!} & 28        & 5.13-11.88 & 75.7 & 2.80 & 2$\cdot 10^{-6}$ & 45.5 & 3 & 1.82 & 3$\cdot$10$^{-3}$& 172.60 & [7] \\ 
           & \texttt{Tiberius}& 26        & 5.13-11.38 & 62.4 & 2.49 & 5$\cdot 10^{-5}$ & 40.6 & 3 & 1.77 & 0.01             & 159.72 & [7] \\ 
           & \texttt{SPARTA}  & 26        & 5.13-11.38 & 56.9 & 2.11 & 7$\cdot 10^{-4}$ & 35.2 & 3 & 1.53 & 0.05             & 180.25 & [7] \\ \hline
WASP-43~b  & \texttt{Eureka!$^{\times0.7}$}& 14 & 5.25-11.75 & 22.9 & 1.76 & 0.04       & 22.6 & 2 & 1.88 & 0.03             & 99.14  & [8] \\ 
           & \texttt{SPARTA}  & 14        & 5.25-11.75 & 24.4 & 1.87 & 0.03             & 21.5 & 2 & 1.79 & 0.04             & 99.75  & [8] \\ \hline
K2-18~b    & \texttt{JExoRES} & 29        & 5.78-11.79 & 29.7 & 1.06 & 0.38             & 29.1 & 2 & 1.08 & 0.36             & 213.08 & [9] \\
           & \texttt{JexoPipe}& 29        & 5.78-11.79 & 32.4 & 1.16 & 0.26             & 29.2 & 3 & 1.12 & 0.30             & 207.97 & [9] \\
           & \texttt{Eureka!} & 27$^a$ & 5.38-11.88 & 54.2 & 2.08 & 1$\cdot 10^{-3}$ & 48.5 & 2 & 1.94 &3$\cdot$10$^{-3}$ & 193.48 & [10]\\
           & \texttt{exoTEDRF}& 27$^a$ & 5.38-11.88 & 53.2 & 2.05 & 1$\cdot 10^{-3}$ & 44.3 & 2 & 1.77 & 0.01             & 195.94 & [10]\\
           & \texttt{SPARTA}  & 27$^a$ & 5.38-11.88 & 71.3 & 2.74 & 4$\cdot 10^{-6}$ & 51.7 & 2 & 2.07 &1$\cdot$10$^{-3}$ & 194.45 & [10]\\ \hline
TOI-270~d  & \texttt{JExoRES} & 26        & 5.49-10.96 & 40.3 & 1.61 & 0.03             & 34.2 & 3 & 1.49 & 0.06             & 199.31 & [11]\\ 
           & \texttt{JexoPipe}& 26        & 5.49-10.96 & 55.7 & 2.39 & 1$\cdot 10^{-4}$ & 50.0 & 3 & 2.43 &1$\cdot$10$^{-1}$ & 187.45 & [11]\\ \hline
TOI-732~c  & \texttt{JExoRES} & 29        & 5.78-11.79 & 44.3 & 1.58 & 0.03             & 37.9 & 3 & 1.46 & 0.06             & 208.61 & [12]\\ 
           & \texttt{JexoPipe}& 29        & 5.78-11.79 & 54.4 & 1.94 & 2$\cdot 10^{-3}$ & 47.6 & 3 & 1.83 &6$\cdot$10$^{-3}$ & 199.97 & [12]\\ \hline            
\end{tabular}
\caption{List of spectra used in the retrieval exploration of sections~\ref{sec:retrieval_setup}-\ref{sec:results}.
$^{\times f}$ Uncertainties multiplied by f (see section~\ref{subsubsec:flat-line-fits}). 
$^a$ First data point removed. $^b$ Last 12 data points removed (see section~\ref{subsubsec:wavelength-ranges}). We give the number of data points N$_p$ and the wavelength range after these removals.
We then give the $\chi^2$ and p values of the flat line fit, and the baseline retrievals (see section~\ref{sec:retrieval_setup}), as well as the ln Z value for the latter.
n$_c$ is the number of constrained parameters, so $\chi^2_{r,base}~=~\frac{\chi^2_{base}}{N_p - n_c}$ (for the flat line fit, n$_c$~=~1). \newline
References: [1] \cite{Kempton2023}; [2] \cite{Zhang_Hu_2024}; [3] \cite{Bouwman2023}; [4] \cite{Alam2025}; [5] \cite{Welbanks_Bell_2024}; [6] \cite{Grant_Lewis_2023}; 
[7] \cite{Powell2024}; [8] \cite{Bell_Crouzet_2024}; [9] \cite{Madhusudhan_Constantinou_2025}; [10] \cite{Luque_Piaulet-Ghorayeb_2025}; [11] \cite{Holmberg_Madhusudhan_2026} ; [12] \cite{Rigby_Madhusudhan_2025a}.
}
\label{tab:spectra-list}
\end{table*}

We first fit each spectrum with a flat line model, to test the null hypothesis of a featureless spectrum. 
This is measured by the best-fit $\chi^2$ (obtained for the mean transit depth weighted by the inverse uncertainties), which we write $\chi^2_{flat}$.
We then convert each $\chi^2_{flat}$ into a p-value, defined as the probability that a flat spectrum with random noise would produce a $\chi^2_{flat}$ higher than the one observed.
If $p \leq 0.05$, then we can reject the null hypothesis of a flat spectrum at 95\% confidence. However the converse is not true, i.e. a spectrum with a higher p (hence a lower $\chi^2_{flat}$) could still have correlated residuals, and would not necessarily imply a flat spectrum.

Apart from GJ~367~b, each planet has at least one spectrum that can reject a flat line, and we demonstrate in section~\ref{subsubsec:pipeline-consistency} that the spectra of the same planet are consistent with each other. 
This justifies searching for atmospheric features fitting these spectra. 
This is specifically interesting for K2-18~b, where we see that the three spectra from \cite{Luque_Piaulet-Ghorayeb_2025} have high $\chi^2_r$ values (1.77-2.07), yielding low p-values ($\leq 10^{-3}$). This means that one could expect to find higher model preference for extra absorbers than in the spectra from \cite{Madhusudhan_Constantinou_2025} that only have $\chi^2_r$ = 1.08 and 1.12, as noticed in \cite{Taylor2025}. 
This is discussed in section~\ref{subsubsec:luque25-spectra}. Finally, we multiply the uncertainties of the \texttt{CASCADE} spectrum of L~168-9~b and the \texttt{Eureka!} (v1) spectrum of WASP-43~b by 0.8 and 0.7 respectively, to have $\chi^2_{r, base} > 1$ for both, so that the retrievals are more likely to find molecules that improve the fit. 
As our conclusion is a non-detection of complex molecules on these spectra (see section~\ref{subsec:results-uninhabitable}), this is a conservative choice. 
All this information is summarized in Table~\ref{tab:spectra-list}.

\begin{figure}
	\includegraphics[width=0.47\textwidth]{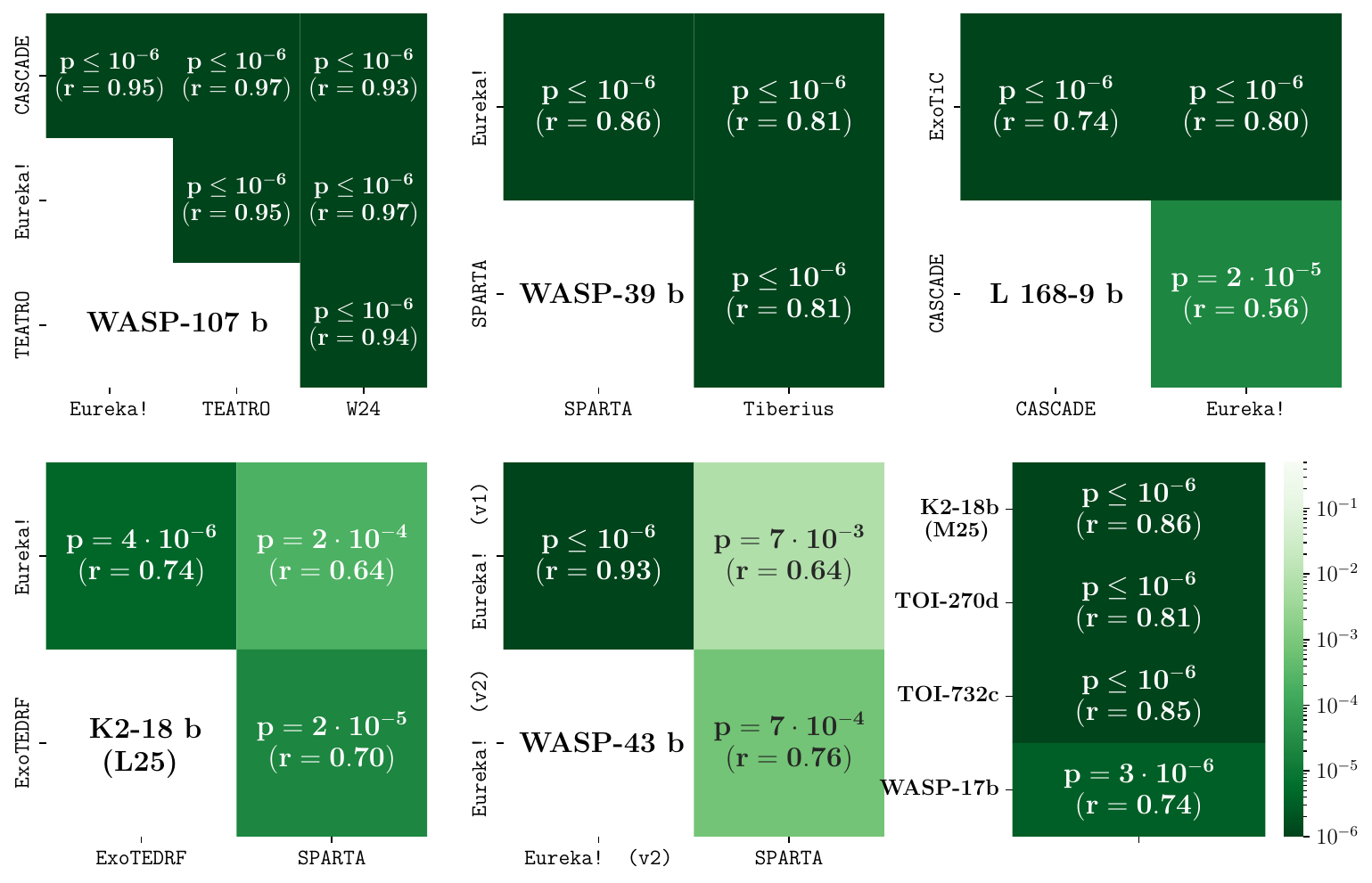}
    \caption{Correlations between different data reductions of the same transmission spectrum observed for each planet (see section~\ref{subsubsec:pipeline-consistency}). 
    We use all the spectra from Table~\ref{tab:spectra-list} and Appendix~\ref{app:extra-spectra}. For each pair of spectra compared, we show the p-value which defines the log-linear color map, and the Pearson coefficient r below. Each of the first five heat maps corresponds to a single planet, and the last one shows the correlations between the only two available reductions of the four planets listed.
    For K2-18~b, the spectra from \cite{Madhusudhan_Constantinou_2025}, shown as M25, cannot be compared with those of \cite{Luque_Piaulet-Ghorayeb_2025}, shown as L25, because of the different wavelength binning.
    }
    \label{fig:correlation-same-planet} 
\end{figure}

\subsubsection{Pipeline Consistency}
\label{subsubsec:pipeline-consistency}
We now attempt to compare multiple spectra of the same planet. We include all planets except GJ~1214~b and GJ~367~b, which only have one spectrum each.
We first remove the mean transit depth (weighted with uncertainties) of each spectrum, then divide the residuals by the uncertainties. 
Then, we compare two resulting arrays X and Y on their overlapping wavelength range with the Pearson correlation coefficient $r_{X,Y}:= \frac{E(XY)-E(X)E(Y)}{\sigma_X \sigma_Y}$. Here, E(X) and $\sigma_X$ are the error-weighted mean (which is zero here by definition) and standard deviations of X respectively. 
$r_{X,Y}$ is always between -1 and 1, with positive and negative values indicating correlation and anticorrelation respectively. Finally, we create samples of random arrays and compute their Pearson coefficients, in order to get a cumulative distribution function and convert our correlation coefficient into p-values, in the same way as for the $\chi^2_{flat}$ metric. 
In this case, p is thus the probability that two random arrays would have a higher correlation coefficient than the one observed. 

We show the results in Figure~\ref{fig:correlation-same-planet}. The \texttt{JExoRES} and \texttt{JexoPipe} spectra of K2-18~b cannot be compared with the other three with this metric, because the wavelength binning is different. We see that all the reduction pipelines give quite consistent spectra for each planet, with all p-values being less than $10^{-2}$, and even $5 \cdot10^{-4}$ except for WASP-43~b. The latter is due to the smaller number of data points.

\begin{figure}
	\includegraphics[width=0.47\textwidth]{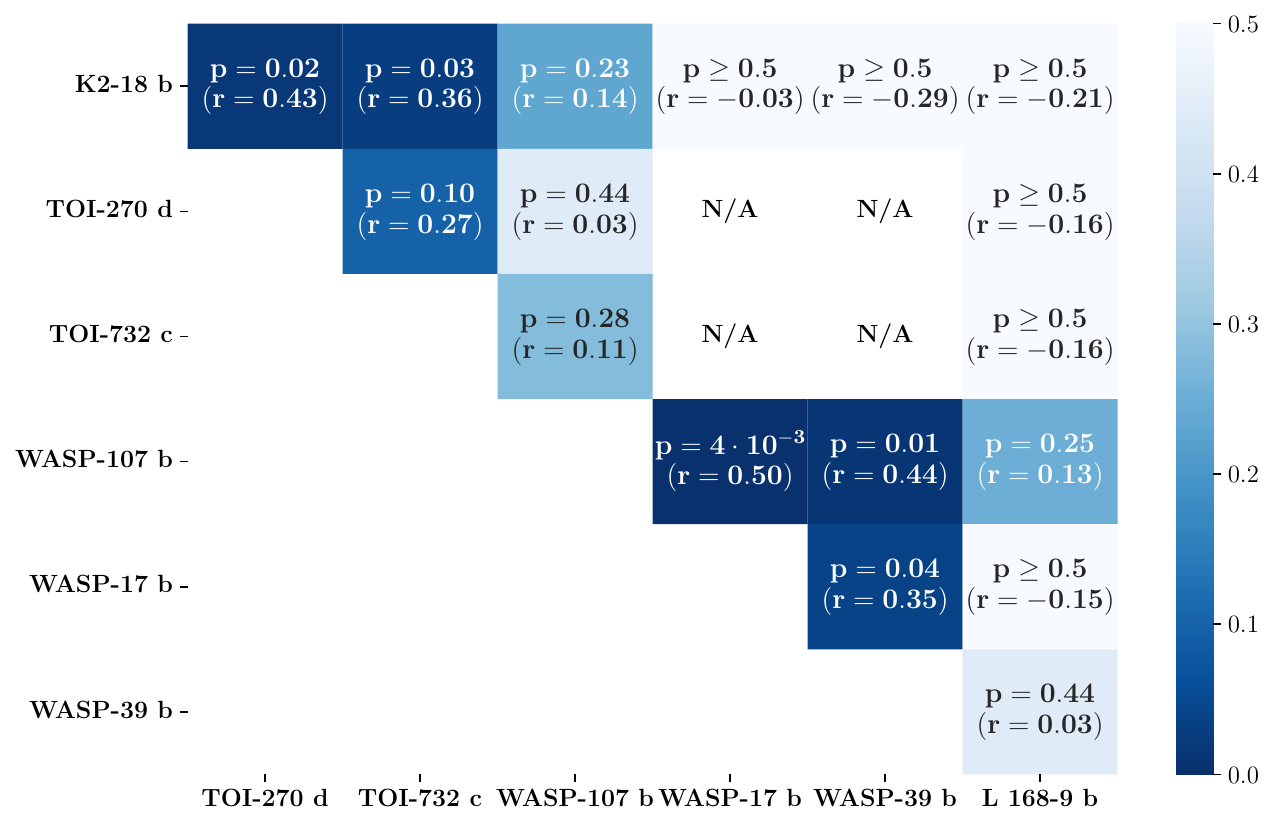}
    \caption{Correlations between transmission spectra of different planets (see section~\ref{subsubsec:planet-correlations}). 
    Note that here the color map is linear between p = 0 to 0.5, because the p-values are higher than in Figure~\ref{fig:correlation-same-planet} (as expected for different planets). TOI-270~d and TOI-732~c cannot be compared with WASP-17~b and WASP-39~b because of the different wavelength binning.
    }
    \label{fig:correlation-different-planets} 
\end{figure}

\subsubsection{Correlations between Planets}
\label{subsubsec:planet-correlations}
Now, we compare the spectra of different planets. This is motivated by the similarities between the three sub-Neptunes in our sample, in terms of size, equilibrium temperature and previously detected species (see Table~\ref{tab:planets-list}). 
We now quantify this using the Pearson correlation coefficient defined in section~\ref{subsubsec:pipeline-consistency}. We compare datasets that have the same wavelength binning, again on their overlaps. 
In addition, we bin down the L~168-9~b and WASP-107~b spectra that have higher resolution, to match the other spectra grids using SpectRes \citep{Carnall_2017}. 
Then, when comparing two planets with $n$ and $m$ different spectra each, there are in total $n \times m$ combinations, so instead of reporting all the values, we take the median correlation coefficients and p-values. 
We show the results in Figure~\ref{fig:correlation-different-planets}. We see that as expected, K2-18~b and TOI-270~d are highly correlated with a median p-value of 0.02. TOI-732~c also shows correlation with K2-18~b (p~=~0.03), and tentative correlation with TOI-270~d (p~=~0.10). 
These three sub-Neptunes do not seem correlated with any other planet (p~$>$~0.2), however WASP-107~b, WASP-17~b and WASP-39~b all seem quite correlated (p~$<$~0.05). 
This is also consistent with the fact that all three are hot gas giants with strong H$_2$O detections (see section~\ref{subsubsec:uninhabitable-planets}). 
This simple analysis already indicates that the features seen with MIRI seem unlikely to be merely random noise (otherwise there would not be such similarities between three planets) or systematics (otherwise these spectra would be quite similar regardless of their type).

\section{Retrieval Setup} 
\label{sec:retrieval_setup}
After this initial analysis of the transmission spectra in our sample, we conduct a more in-depth analysis using atmospheric retrievals \citep{Madhusudhan2009, benneke2013, Madhusudhan2018}, with the open-source retrieval code \texttt{petitRADTRANS} (\texttt{pRT}) \citep{Molliere2019, Nasedkin2024}. 
In each of these spectra, we look for multiple potential trace species, following the approach of \cite{Madhusudhan_Constantinou_2025}, \cite{Rigby_Madhusudhan_2025a}, \cite{Pica-Ciamarra_Madhusudhan_2026}, \cite{Welbanks_Nixon_2026} and \cite{Holmberg_Madhusudhan_2026}. Our main goal is to determine whether or not the additional absorbers found in these works are also found in other planets.
To do this, we run retrievals with a selection of N baseline species depending on the planet, and one additional molecule X at a time (we refer to these as \textquote{N+X} retrievals). We compare the Bayesian evidence of each retrieval with one containing only the baseline species (which we refer to as the baseline retrieval), to get a model preference $\ln B$ for each molecule X for the given spectrum. We call $\ln B_{max}$ and $\ln B_{min}$ the highest and lowest preference obtained for a molecule across all the spectra of a given planet.

In what follows, we describe the general framework of the retrieval model in section~\ref{subsec:generic-info}. We then motivate our choices of baseline species in section~\ref{subsec:baseline-species}, and discuss the opacity sources explored in section~\ref{subsec:species-explored}. 
Then, we describe the temperature and cloud parametrization in section~\ref{subsec:temp-and-clouds}, and the bulk parameters fitting in section~\ref{subsec:bulk-parameters}.
Finally, we specify the different treatment of GJ~1214~b, L~168-9~b and GJ~367~b in section~\ref{subsec:exception-gj-and-l168}, and of WASP-107~b in section~\ref{subsec:exception-WASP-107b}. All the parameter priors are summarized in Appendix~\ref{app:retrieval_priors}.

\subsection{General Framework}
\label{subsec:generic-info}
The retrieval framework considers the terminator atmosphere in one dimension, assuming hydrostatic equilibrium and uniform molecular abundances with pressure. The model has 100 pressure layers between 10$^{-6}$ and 10 bar, spaced log-uniformly. We include collision-induced absorption for H$_2$-H$_2$, H$_2$-He and He-He \citep{borysow1988, orton2007, abel2011, richard_new_2012}. We sample from the posterior probability distribution using \texttt{MultiNest} \citep{Feroz2009}. We use 500 live points for most retrievals, but we increase this to 700 for WASP-107~b because of the higher SNR, and to 1000 for each baseline retrieval. The latter improves the precision of all Bayesian comparisons, with minimal increase in computing time.

\subsection{Choice of Baseline Species}
\label{subsec:baseline-species}
In all our retrievals, we assume H$_2$ and He (He/H~=~0.17 by number) as the background gas and fit for the abundances of additional species. This is because most of the planets were found to be H$_2$/He dominated (see section~\ref{subsec:planets}). 
The exceptions to this are GJ~1214~b, L~168-9~b and GJ~367~b; we explain their different treatments in section~\ref{subsec:exception-gj-and-l168}.Apart from H$_2$ and He, in the baseline retrievals, we add the species listed in Appendix~\ref{app:retrieval_priors}. For the three sub-Neptunes, these are chosen to be consistent with previous studies: CH$_4$+CO$_2$ for K2-18~b (\citealt{Madhusudhan_Constantinou_2025, Pica-Ciamarra_Madhusudhan_2026}, 2+X setup), CH$_4$+CO$_2$+H$_2$O for TOI-270~d (\citealt{Holmberg_Madhusudhan_2026}, 3+X), and CH$_4$+CO$_2$+H$_2$O+CO for TOI-732~c (\citealt{Rigby_Madhusudhan_2025a}, 4+X). 
For WASP-39~b, WASP-17~b and WASP-43~b, we only add H$_2$O (1+X setup), which was found with the highest significance in previous MIRI works (see section~\ref{subsec:planets}). 
In general, the effect a simpler baseline retrieval, is to increase the evidence for additional species. This was shown in previous works including \cite{Madhusudhan_Constantinou_2025}, and in this work in Appendix~\ref{app:WASP-39b-SO2}. As mentioned in section~\ref{subsubsec:flat-line-fits} and shown in section~\ref{subsec:results-uninhabitable}, our conclusion in this work is a non-detection of complex molecules in these spectra, so this choice is conservative. Finally, we add SO$_2$+H$_2$O+H$_2$S+PH$_3$+SO for WASP-107~b (5+X), and discuss this choice in section~\ref{subsec:exception-WASP-107b}.

One difference of \texttt{pRT} from most retrieval codes is that the abundance priors for species other than H$_2$/He are defined in mass fractions rather than volume mixing ratios (VMRs). In order to approximately match the standard bounds in the field for log(VMR) of 10$^{-12}$ to 10$^{-0.3}$, we choose mass fraction bounds of  10$^{-11}$ to 1, except for WASP-107~b (see section~\ref{subsec:exception-WASP-107b}).

\subsection{Species Explored} 
\label{subsec:species-explored}
As mentioned earlier, we search for one molecule X at a time, in addition to the baseline ones in Appendix~\ref{app:retrieval_priors}. In total, we search for 160 such molecules, listed in Appendix~\ref{app:species-list}.
Since \texttt{pRT} implements correlated-k opacities natively, we use molecular opacities in the correlated-k mode \citep{Lacis_Oinas_1991} natively available in \texttt{pRT} or obtained from the ExoMol opacity database \citep{Chubb_Rocchetto_2021, Tennyson_Yurchenko_2024} when available (CH$_3$Cl, CH$_3$F, CH$_4$, CO, CO$_2$, CS, CS$_2$, C$_2$H$_2$, C$_2$H$_4$, HCN, H$_2$CS, H$_2$O, H$_2$S, NH$_3$, OCS, PH$_3$, SiH$_4$, SiO, SO, SO$_2$, SO$_3$).

For the remaining species where only HITRAN cross sections are available \citep{Gordon_Rothman_2026}, we use the data at $T\approx300$~K and $P \approx 1$~bar, or the closest available temperature and pressure, and we convert them to the correlated-k format. The exception is ethane (C$_2$H$_6$) for which we take the HITRAN line list and convert it into a \texttt{pRT} correlated-k table, using Cthulhu \citep{Agrawal_MacDonald_2024} and Exo\_k \citep{Leconte_2021}. We use the \texttt{pRT} default model resolution R = 1000, and number of Gaussian points N$_g$ = 16.

\subsection{Temperature Profile and Clouds}
\label{subsec:temp-and-clouds}
In our retrievals, we apply an isothermal P-T profile with uniform priors depending on the planets' equilibrium temperatures (see Appendix~\ref{app:retrieval_priors}). 
In Appendix~\ref{app:sensitivity-bulk-priors}, we show that the results are mostly insensitive to setting narrower temperature priors.
This reduces computation time. We also verified that for most planets, a more complex P-T profile prior yielded a temperature posterior consistent with an isotherm.
For aerosols, we use a patchy cloud deck, with only two parameters: the opaque cloud top pressure $P_c$, and the cloud coverage fraction $\phi$. This corresponds to two of the four parameters in the cloud+haze setup of \cite{macdonald2017, pinhas2019}, because the haze in that prescription  only provides significant extinction in the visible/NIR.

\subsection{Bulk Parameters}
\label{subsec:bulk-parameters}
For each planet, we fix its radius, mass, and host star radius according to the values in Table~\ref{tab:planets-list}, and fit the reference pressure between 10$^{-6}$ and 10 bar. 
In reality, the masses found from radial velocity observations have uncertainties. 
In Appendix~\ref{app:sensitivity-bulk-priors}, we show that our results are mostly insensitive to adding Gaussian priors on the planetary mass, or replacing the reference pressure prior by a radius prior.

\subsection{Exception 1: Featureless Spectra}
\label{subsec:exception-gj-and-l168}
As mentioned in sections \ref{subsec:planets} and \ref{subsec:baseline-species}, GJ~1214~b, L~168-9~b and GJ~367~b have shown mostly featureless JWST MIRI-LRS spectra \citep{Kempton2023, Bouwman2023, Zhang_Hu_2024}, indicative of either high clouds or non-H$_2$/He dominated atmospheres. This means that in order to constrain their atmospheric composition, one could use different abundance priors to the ones described in section~\ref{subsec:baseline-species}, for instance central-log-ratio (CLR) priors \citep{Benneke_Seager_2012}. However, because our goal is to conduct a comparative analysis of the ten exoplanet spectra, we only make small changes to the setup described in section~\ref{subsec:baseline-species}. First, we choose CO$_2$ as a baseline species, because it is a very weak absorber in the $5-12 \mu$m range (see, for instance, Figure 2 of \citealt{Pica-Ciamarra_Madhusudhan_2026}), so its main purpose is to allow the mean molecular weight (MMW)  to increase. We also restrict its mass fraction prior between 10$^{-3}$ and 1, because any value below $\approx10^{-2}$ will have a negligible effect on the MMW. Finally, we restrict the mass fraction of the extra molecule added between 10$^{-11}$ and 10$^{-1}$, so that it does not replace H$_2$/He or CO$_2$ as the main gas. We note that for these three planets, we tested replacing CO$_2$ by H$_2$O or combining both as baseline species, which did not have a considerable impact on the results presented in section~\ref{subsec:results-uninhabitable}.

\subsection{Exception 2: WASP-107~b}
\label{subsec:exception-WASP-107b}
As mentioned earlier, the WASP-107~b spectra have by far the highest SNR. This is evident from the high $\chi_{r,flat}^2$ metric (see section~\ref{subsubsec:flat-line-fits} and Appendix~\ref{app:extra-spectra}).
First, this means that we need to add multiple molecules in the baseline retrieval for it to achieve a reasonable fit. We first try with H$_2$O, SO$_2$ and H$_2$S, the three gas-phase species found with highest significance in \cite{Dyrek_Min_2024}. With each spectrum, these three species are well constrained, but the $\chi_{r}^2$ are still too high, ranging from 3 to 6, whereas all the other planets are below 2.5. 
The lowest value of 3 is obtained with the \texttt{W24} spectrum, so we choose to focus on this one in what follows. Also, we notice that the retrievals are pushing toward high mass fractions of H$_2$S in order to increase the MMW. However, WASP-107~b being a gas giant, its atmosphere must be H$_2$/He dominated. So as in section~\ref{subsec:exception-gj-and-l168}, but here for all the molecules except H$_2$/He, we restrict the mass fraction priors between 10$^{-11}$ and 10$^{-1}$.

The next step to find a well-suited baseline set of molecules, is that we run a large retrieval with all the same gas-phase species as in \cite{Dyrek_Min_2024}: H$_2$O, SO$_2$, H$_2$S, NH$_3$, CO, PH$_3$, HCN, C$_2$H$_2$, SiO, CH$_4$, CO$_2$ and SO. We keep basic patchy clouds rather than silicate clouds, and we note (as mentioned in section~\ref{subsubsec:uninhabitable-planets}) that \cite{Dyrek_Min_2024} used the MIRI data together with HST data, so we do not expect the exact same results. 
We find strong constraints on H$_2$O, SO$_2$, H$_2$S, PH$_3$ and SO, a weak constraint on SiO, and upper bounds on the other species. We then run a new retrieval with just the first five, and get $\chi_{r}^2 = 2.02$, which is a comparable value to other baseline retrievals, so we choose this as the WASP-107~b baseline.

Finally, the high SNR and the five well constrained molecules in the baseline cause the retrievals to be computationally expensive, so we decide to only use the \texttt{W24} spectrum.
In addition, we only explore a limited selection of 44 molecules, given in Appendix~\ref{app:species-list}. It includes the 19 that are favored in other planets, listed in Table~\ref{tab:summary-molecules}.

\begin{figure}
	\includegraphics[width=0.473\textwidth]{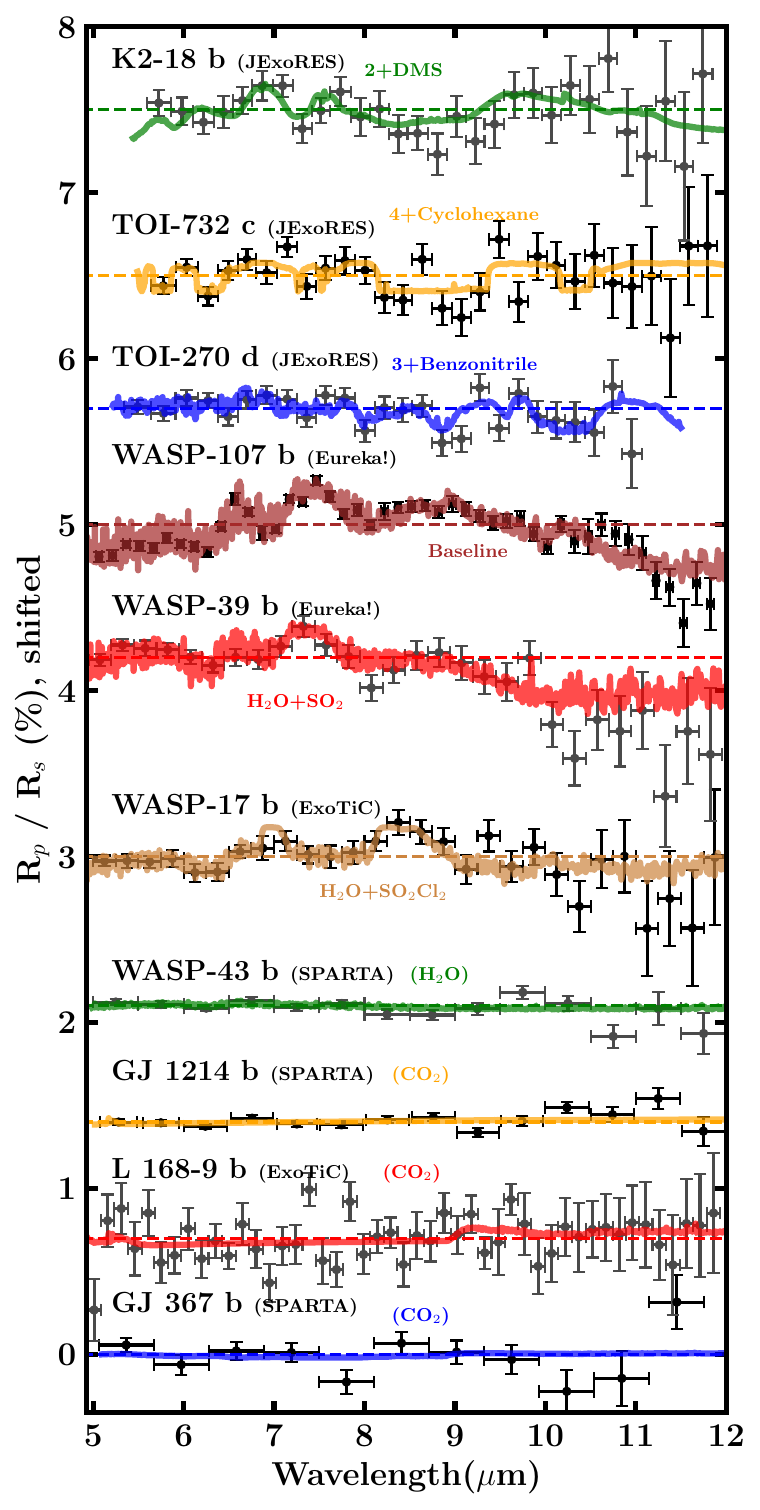}
    \caption{JWST MIRI-LRS transmission spectra of all 10 planets considered in this work. In black and gray are the data, from the indicated reduction pipeline. In colored dashed lines are the mean transit depths, and in colored solid lines are the best-fit model spectra from the retrieval including the molecule reaching the highest $\ln B_{min}$ for this planet, if $\ln B_{min} \geq 2$  (see section~\ref{sec:results}). When no molecule reaches this threshold, we display the baseline retrieval spectra, which are discussed in section~\ref{subsec:baseline-species} and summarized in Appendix~\ref{app:retrieval_priors}. We write (H$_2$O) or (CO$_2$) when the molecule is weakly constrained.
    Each spectrum has been converted from transit depth $(\frac{R_p}{R_\star})^2$ into radius ratio $\frac{R_p}{R_\star}$, as in Appendix~\ref{app:chi2-comparison}.
    }
    \label{fig:all-spectra} 
\end{figure}

\section{Results}
\label{sec:results}
In this section, we present the results of our species exploration, described in section~\ref{sec:retrieval_setup}.
As in \cite{Rigby_Madhusudhan_2025a}, \cite{Pica-Ciamarra_Madhusudhan_2026} and \cite{Holmberg_Madhusudhan_2026}, we consider that a molecule is a good candidate to explain the planet's extra absorption if it reaches $\ln B \geq 2$ across all data reductions used, in other words $\ln B_{min}$ must be $\geq 2$. For completeness, we will also report some molecules that are below this threshold, although we do not consider them to be good candidates.
We decide to not convert the $\ln B$ values into \textquote{N$\sigma$} detection significances, following the recommendation of \cite{Thorngren_Sing_2026}, although we show a conversion table in Appendix~\ref{app:sigma-conversion}. This previous work also recommended using the BPICS metric in addition to the Bayes factor, so we do this and show the results in Appendix~\ref{app:BPICS}.

In Figure~\ref{fig:all-spectra}, we plot one spectrum for each planet, along with its best-fit model for the molecule that achieved the highest $\ln B_{min}$. Each spectrum has been converted from transit depth $(\frac{R_p}{R_\star})^2$ into radius ratio $\frac{R_p}{R_\star}$, so that the scatter and uncertainties are of comparable size. These converted spectra are also the ones used in the $\chi^2$ comparison presented in Appendix~\ref{app:chi2-comparison}.
We then show more precise comparisons of multiple retrievals for each planet, in Appendix~\ref{app:spectral-fits}. 

\subsection{Uninhabitable Planets}
\label{subsec:results-uninhabitable}
We start with presenting the retrieval results for six of the seven uninhabitable planets of our sample, excluding WASP-107~b which we show in section~\ref{subsec:WASP-107b-results}. We first report the constraints obtained from the baseline retrievals, including H$_2$O abundance and temperature for WASP-39~b and WASP-17~b. Then, we list the additional absorbers that are potentially inferred in these planets.

\subsubsection{Baseline Fits}
\label{subsubsec:baseline-fits}
These six planets are GJ~1214~b, L~168-9~b, WASP-17~b, WASP-39~b and WASP-43~b. As explained in section~\ref{subsec:baseline-species}, their baseline retrievals only include H$_2$/He and H$_2$O or CO$_2$. We find H$_2$O contributions in WASP-17~b and WASP-39~b, consistent with \cite{Grant_Lewis_2023} and \cite{Powell2024} respectively. In contrast, we do not find any H$_2$O contribution for WASP-43~b unlike in its previous phase curve study \citep{Bell_Crouzet_2024}. 
However, this is not necessarily contradictory, as the transit alone offers less information content. 
Finally, we also do not find any CO$_2$ contribution in GJ~1214~b, L~168-9~b or GJ~367~b, as expected from the spectra being flat and CO$_2$ absorbing very weakly in this range.
So, for these four planets, the atmospheric model is just marginally a better fit than a flat line, as shown in Table~\ref{tab:spectra-list}. Thus, the atmospheric temperature is only well-constrained for WASP-17~b and WASP-39~b at T~$\approx1550-1900$~K and $1050-1200$~K respectively (depending on the data reduction), which is consistent with their respective $T_{eq}$ values of $\sim$  1750 K and 1150 K. On the other hand, the patchy cloud deck stays unconstrained. Having not included SO$_2$ for WASP-39~b, the baseline fits are underfitting ($\chi^2_{base} \geq 1.5$), unlike in WASP-17~b where $\chi^2_{base} \leq 1.2$, so we expect more potential extra absorbers to be inferred in the former (see the following section~\ref{subsubsec:uninhabitable-exploration-result}). 

\subsubsection{Additional Absorbers}
\label{subsubsec:uninhabitable-exploration-result}
We now present the exploration results for these uninhabitable planets, summarized in Table~\ref{tab:exploration-uninhabitable}. As discussed in section~\ref{subsec:baseline-species}, for these planets we consider the 1+X exploration with H$_2$O or CO$_2$ being the only baseline species.
Out of the $\geq 150$ species explored, we find that only two species have $\ln B_{min} \geq 2$ for at least one planet: SO$_2$ in WASP-39~b and SO$_2$Cl$_2$ in WASP-17~b.
The former was expected, because it was also found in \citep{Powell2024}. We show in Appendix~\ref{app:P24-consitency} that our results are consistent with this previous work. 
On the other hand, WASP-17~b's extra absorption at $\lambda \approx 8.5-9 \mu m$ is more likely to be caused by silicate clouds \citep{Grant_Lewis_2023} then by SO$_2$Cl$_2$, a chemical that has never been found naturally on Earth nor detected in space.

We find a few other species with weaker evidence of $\ln B_{max} \geq 2$. In this case, we infer 6 other molecules (CH$_4$, SO$_3$, SOF$_2$, C$_2$H$_2$, CH$_3$Cl and HCN) in WASP-39~b, and benzene (C$_6$H$_6$) in WASP-43~b. For WASP-39~b, we show in Appendix~\ref{app:WASP-39b-SO2} that this model preference vanishes, (i.e. falls below $\ln B_{max}=1$), for all of them when including SO$_2$ in the baseline retrieval. As for benzene, it could potentially be present in WASP-43~b's atmosphere if there were also lower-order hydrocarbons like CH$_4$ and C$_2$H$_6$, however we would not expect it at the retrieved abundances of 10 ppm to 1\% \citep{Moses_Fouchet_2005, Nixon_2024}.

\begin{table*}
\renewcommand{\arraystretch}{1.05} 
\setlength{\tabcolsep}{3pt}
\centering
    \begin{tabular}{l|cccccc|ccc} \hline \hline 
        Molecule  & WASP-39~b   & WASP-17~b & WASP-43~b & GJ~1214~b & L~168-9~b & GJ 367~b & K2-18~b & TOI-270~d & TOI-732~c \\ \hline
       SO$_2$ &$\mathbf{\und{3.1-6.3}^{max}}$ &$-$     & $-$           & $-$ & $<0.9 $ & $-$  & $-$     & $-$    & $-$     \\
 SO$_2$Cl$_2$& $-$  &$\mathbf{\und{2.8-3.0}^{max}}$   &$-$             & $-$ & $-$   &$-$   & $<0.5$  & $-$    & $-$    \\ \hline
        CH$_4$       & \und{0.8$-$5.3} & $<0.1$    & $-$              & $-$ & $-$  &$-$    & $<1.0$  & $<0.3$ & $<0.1$  \\
        SO$_3$       & 1.3$-$3.4 & $-$       & $<0.3 $          & $-$ & $-$   &$-$   & $<0.6$  & $-$    & $<1.2$  \\
        SOF$_2$      & 1.5$-$2.9 & $-$       & $-$              & $-$ & \und{$<1.0$} &$-$  & $-$     & $-$    & $-$     \\
        C$_2$H$_2$   & 0.2$-$2.2 & $-$       & $-$              & $-$ & $-$  &$-$    & $<1.3$  & $<0.1$ & $<0.5$  \\
        CH$_3$Cl     & 0.2$-$2.1 & $-$       & $<0.1$           & $-$ & $-$  &$-$    & 0.4$-$3.0& $<0.1$ & $<1.5$  \\
        HCN          & 0.1$-$2.0 & $-$       & $-$              & $-$ & $-$  &$-$    & $<1.3$  & $<0.3$ & $<1.9$  \\
        Benzene      & $-$        & $-$&\und{0.1$-$2.7}$^{max}$&$-$ & $-$  & 0.3 &  $<1.0$  & $-$  & 1.4$-$4.4\\ 
        
        Cyclohexane  & $<0.1$     & $-$       & \und{$<1.2$}           & $-$ & $-$  &$-$     &-1.1$-$2.1& $-$   & $\mathbf{\und{4.3-6.2}^{max}}$\\ 
        C$_2$H$_4$ & $-$        & $-$       & $-$      & \und{1.6}$^{max}$ & $-$  &$0.2$     &$<1.3$&$-$ & $<1.7$\\
        Cyclopentane & $-$        & $-$       & $-$      & \und{1.5}         & $-$  &$-$     &-1.0$-$2.7&$<0.1$ & 0.9$-$3.1\\
        CCl$_2$S     & $<0.7$     & $-$       & $-$      & $-$&$\und{<1.3}^{max}$&$-$ &$-$      & $-$   & $-$\\
        Toluene         & $<0.4$     & $-$       & $<1.3$   & $-$    &$-$     & \und{0.7$^{max}$}   & $<0.5$     & $<0.9$   & $<0.8$\\
        \hline
    \end{tabular} 
    \vspace{2mm}
    \caption{Model preferences for the 2 species that reach $\ln B_{min} \geq 2$ in uninhabitable planets, and the 12 additional species that are highlighted in section~\ref{subsubsec:uninhabitable-exploration-result}. If $\ln B_{max} \geq 2$, we write \textquote{$\ln B_{min}-\ln B_{max}$}; if $0<\ln B_{max}<2$, we write \textquote{$< \ln B_{max}$}; finally if $\ln B_{max}\leq0$ we write an ellipsis. \Bf{Bold} indicates $\ln B_{min} \geq 2$. 
    \und{Underlined}: Spectrum plotted in Appendix~\ref{app:spectral-fits}.
    Superscript \textquote{max} indicates the species with the highest $\ln B_{max}$ for this planet. 
    }
    \label{tab:exploration-uninhabitable}
\end{table*}

Next, 5 additional species are worth highlighting, even though they have $\ln B_{max} < 2$.
First, C$_2$H$_4$, CCl$_2$S and toluene (C$_7$H$_8$) are the molecules with the highest $\ln B_{max}$ values in GJ~1214~b (1.6), L~168-9~b (1.3) and GJ~367~b (0.7)  respectively. 
We also mention cyclopentane (C$_5$H$_{10}$) which obtains $\ln B = 1.5$ in the GJ~1214~b spectrum, because it is the only molecule apart from SO$_2$ and SO$_2$Cl$_2$ that crosses our threshold with the BPICS metric for one of these six planets (see Appendix~\ref{app:BPICS}).
Finally, we note that cyclohexane (C$_6$H$_{12}$) reaches $\ln B_{max} = 1.2$ in WASP-43~b, and is inferred in two of the temperate sub-Neptunes, namely K2-18~b (only in the \texttt{JexoPipe} spectrum, at $\ln B = 2.1$) and TOI-732~c. In the latter, it is the molecule with the highest model preference for both the \texttt{JExoRES} ($\ln B = 4.3$) and \texttt{JexoPipe} ($\ln B = 6.2$) spectra (more information on this in section~\ref{subsec:sub-Neptune-results}). 

\subsection{Temperate Sub-Neptunes}
\label{subsec:sub-Neptune-results}
Now, we present the results for the three temperate sub-Neptunes. We expect to see similarities between their retrieval results, because of the spectral similarities presented in section~\ref{subsubsec:planet-correlations}.

\subsubsection{Re-analysis of Previous Spectra}
\label{subsubsec:previous-sub-Neptunes}
We start by presenting the exploration results for the \texttt{JExoRES} and \texttt{JexoPipe} spectra of K2-18~b and TOI-732~c. These were previously run with the retrieval code \texttt{POSEIDON} \citep{macdonald2017, macdonald_poseidon_2024}, in \cite{Pica-Ciamarra_Madhusudhan_2026} and \cite{Rigby_Madhusudhan_2025a} respectively. 
For K2-18~b, we run CH$_4$+CO$_2$+X (i.e. 2+X) retrievals and find 14 species with $\ln B \geq 2$ for the \texttt{JExoRES} spectrum, and 17 for the \texttt{JexoPipe} spectrum. We find that 10 of these species are above the threshold for both spectra, which is consistent with the finding of  \cite{Pica-Ciamarra_Madhusudhan_2026}, and we show them in Table~\ref{tab:K2-18b-M25-results}. The only difference is that we do not find cyclohexane (C$_6$H$_{12}$), but instead find diethyl sulfide (C$_4$H$_{10}$S). We also confirm that the species with the highest preference is chloroethane (C$_2$H$_5$Cl, $\ln B = 3.2$ and $4.4$ in \texttt{JExoRES} and \texttt{JexoPipe} respectively).

\begin{table}
\renewcommand{\arraystretch}{1.05} 
\setlength{\tabcolsep}{4.5pt}
\centering
    \begin{tabular}{l|cc} \hline \hline 
        Molecule                         & $\ln B_{JRES}$ & $\ln B_{JPipe}$ \\ \hline
        Chloroethane (C$_2$H$_5$Cl)      & 3.2            & 4.4             \\   
        Methacrylonitrile (C$_4$H$_5$N)  & 2.6            & 2.9             \\ 
        Butane (C$_4$H$_{10}$)           & 2.6            & 2.6             \\ 
        Propyne (C$_3$H$_4$)             & 2.5            & 2.8             \\  
        DMS (C$_2$H$_6$S)   & 2.2            & 2.8             \\ 
        Diethyl sulfide (C$_4$H$_{10}$S) & 2.1            & 2.7             \\ 
        Allyl chloride (C$_3$H$_5$Cl)    & 2.1            & 2.3             \\ 
        Bromoethane (C$_2$H$_5$Br)       & 2.5            & 2.0             \\             
        Dichloromethane (CH$_2$Cl$_2$)   & 2.0            & 2.6             \\    
        Cyclopentane (C$_5$H$_{10}$)     & 2.0            & 2.1             \\ \hline
    \end{tabular} 
    \vspace{2mm}
    \caption{K2-18~b M25 2+X results (see section~\ref{subsubsec:previous-sub-Neptunes}). $\ln B_{JRES}$ and $\ln B_{JPipe}$ are the evidence obtained with the \texttt{JExoRES} and \texttt{JexoPipe} spectra respectively. We show the 10 species with $\ln B_{min,M25} \geq 2$.
    These values are consistent with Table~1 of \cite{Pica-Ciamarra_Madhusudhan_2026}.
    }
    \label{tab:K2-18b-M25-results}
\end{table}

For TOI-732~c, we run CH$_4$+CO$_2$+H$_2$O+CO+X (i.e. 4+X) retrievals and also find 10 species with $\ln B \geq 2$ across both spectra, which we list in Table~\ref{tab:TOI-732c-results}. Only two are common with K2-18~b: chloroethane and methacrylonitrile (C$_4$H$_5$N). These were not highlighted in \cite{Rigby_Madhusudhan_2025a}, because chloroethane was not considered, and the method only looked in the MIRI data for species that had $\ln B\geq 2$ in the NIR data, which was not the case for methacrylonitrile. As mentioned in section~\ref{subsubsec:uninhabitable-exploration-result}, the highest model preference is reached for cyclohexane, at $\ln B = 4.3$ and $6.2$ in \texttt{JExoRES} and \texttt{JexoPipe}, respectively. We also highlight benzene, which was found in one of the WASP-43~b spectra, and is favored at $\ln B = 4.4$ in the \texttt{JexoPipe} spectrum of TOI-732~c (although only 1.4 in the \texttt{JExoRES} spectrum). Lastly, we also look at ethane because it would be expected to be present in the atmosphere if there are also higher-order hydrocarbons \citep{Moses_Fouchet_2005, Nixon_2024}. We find $\ln B = 4.3$ in the \texttt{JexoPipe} spectrum, but only $\ln B = 1.0$ in the \texttt{JExoRES} one. 

Finally, for TOI-270~d, the exploration was conducted with the exact same \texttt{pRT} setup as in \cite{Holmberg_Madhusudhan_2026}, so we only need to run a few species that were not considered previously. We note that this previous work had more than 200 species, so the selection was reduced for this current work, however we made sure to include all the species that reached $\ln B \geq 2$ in at least one spectrum. We show the results in Table~\ref{tab:TOI-270d-results}. In this planet, only three molecules have $\ln B_{min} \geq 2$: benzonitrile (C$_7$H$_5$N), methacrylonitrile and isobutene (C$_4$H$_8$). These were already found in \cite{Holmberg_Madhusudhan_2026}, and isobutene was highlighted for both, being favored in TOI-270~d, and being one of only two species favored across all near- and mid-infrared datasets of TOI-732~c \citep{Rigby_Madhusudhan_2025a}. We further highlight methacrylonitrile, for being favored in all six spectra of temperate sub-Neptunes considered so far. 

\begin{table}
\renewcommand{\arraystretch}{1.05} 
\setlength{\tabcolsep}{4.5pt}
\centering
    \begin{tabular}{l|cc} \hline \hline 
        Molecule                        & $\ln B_{JRES}$ & $\ln B_{JPipe}$ \\ \hline
        Cyclohexane (C$_6$H$_{12}$)     & 4.3            & 6.2             \\   
        3-Methylpentane (C$_6$H$_{14}$) & 3.7            & 4.3             \\ 
        2-Butene (C$_4$H$_8$)           & 3.3            & 5.0             \\ 
        3-Methylhexane (C$_7$H$_{16}$)  & 2.8            & 4.0             \\  
        1-Pentene (C$_5$H$_{10}$)       & 2.4            & 2.5             \\ 
        Methacrylonitrile (C$_4$H$_5$N) & 2.4            & 4.3             \\ 
        Chloroethane (C$_2$H$_5$Cl)     & 2.4            & 2.4             \\ 
        Isobutene (C$_4$H$_8$)          & 2.3            & 4.8             \\             
        cis-2-Pentene (C$_5$H$_{10}$)   & 2.1            & 2.7             \\    
        Hexane (C$_6$H$_{14}$)          & 2.0            & 5.8             \\ \hline
    \end{tabular} 
    \vspace{2mm}
    \caption{TOI-732~c 4+X results (see section~\ref{subsubsec:previous-sub-Neptunes}). We show the 10 species with $\ln B_{min} \geq 2$.
    Out of these, only 1-pentene and isobutene were also looked for in the MIRI data in \cite{Rigby_Madhusudhan_2025a} (Table 2), and these results are consistent.
    }
    \label{tab:TOI-732c-results}
\end{table}

\begin{table}
\renewcommand{\arraystretch}{1.05} 
\setlength{\tabcolsep}{3pt}
\centering
    \begin{tabular}{l|cc} \hline \hline 
        Molecule                         & $\ln B_{JRES}$ & $\ln B_{JPipe}$ \\ \hline
        Benzonitrile (C$_7$H$_5$N)       & 3.4            & 5.3             \\
        Methacrylonitrile (C$_4$H$_5$N)  & 2.5            & 3.2             \\
        Isobutene (C$_4$H$_8$)           & 2.3            & 4.3             \\ \hline
    \end{tabular} 
    \vspace{2mm}
    \caption{TOI-270~d 3+X results from \cite{Holmberg_Madhusudhan_2026}. Only 3 species have $\ln B_{min} \geq 2$.
    }
    \label{tab:TOI-270d-results}
\end{table}

\subsubsection{Analysis of New K2-18~b Spectra}
\label{subsubsec:luque25-spectra}
In addition to these six spectra, we also perform our retrieval exploration on the three K2-18~b spectra obtained with different data reductions in \cite{Luque_Piaulet-Ghorayeb_2025}, without the first data point (see section~\ref{subsubsec:wavelength-ranges}). We show the results in Table~\ref{tab:K2-18b-L25-results}. As predicted in section~\ref{subsubsec:flat-line-fits}, the higher $\chi_{flat}^2$ values of these spectra allow for some species to reach much higher model preferences than with the \texttt{JExoRES} and \texttt{JexoPipe} spectra of K2-18~b. Namely, the \texttt{exoTEDRF} and \texttt{SPARTA} spectra are best fitted by ethyl mercaptan (C$_2$H$_6$S, an isomer of DMS)  at $\ln B = 6.3$ and $6.1$ respectively, and the \texttt{Eureka!} spectrum by 2-butene at a surprisingly high $\ln B = 9.7$. We also find that a lot more molecules reach the $\ln B \geq 2$ threshold: 57, 34 and 28 for the \texttt{Eureka!}, \texttt{exoTEDRF} and \texttt{SPARTA} data respectively. 19 of these species are common between the three spectra, with ethyl mercaptan the only one at $\ln B \geq 5$ in all three.
Finally, out of these 19 species, only 3 are in common with the \texttt{JExoRES} and \texttt{JexoPipe} list: DMS ($\ln B = 2.2-5.6$), diethyl sulfide (C$_4$H$_{10}$S, $\ln B = 2.1-9.1$) and chloroethane ($\ln B = 2.1-8.0$). Out of these three, only chloroethane is found in both spectra of another temperate sub-Neptune, TOI-732~c. We also highlight that we find $\ln B = 3.0$ for CH$_3$Cl with the \texttt{Eureka!} spectrum, which was also found at $\ln B = 2.1$ in the \texttt{Tiberius} spectrum of WASP-39~b (see section~\ref{subsubsec:uninhabitable-exploration-result}). In both cases, it is not robust to the different data reductions. Finally, we mention ethane, which was argued to be degenerate with DMS in \cite{Luque_Piaulet-Ghorayeb_2025}. We do not find ethane at $\ln B \geq 2$ in any of the five K2-18~b spectra, but it does reach a value of $\ln B = 1.8$ with the \texttt{Eureka!} reduction. 

We note that just like for most of the uninhabitable planet data in our sample, these MIRI spectra were never studied alone, but only together with the NIR data. So, our results are not directly comparable with those of \cite{Luque_Piaulet-Ghorayeb_2025}. 

\begin{table}
\renewcommand{\arraystretch}{1.05} 
\setlength{\tabcolsep}{3pt}
\centering
    \begin{tabular}{l|cc} \hline \hline 
        Molecule                                & $\ln B_{min,L25}$ & $\ln B_{min,M25}$ \\ \hline
        Ethyl mercaptan (C$_2$H$_6$S)           & 6.1               & 1.6               \\   
        Cyclooctane (C$_8$H$_{16}$)             & 4.2               & 0.0               \\ 
        2-Butene (C$_4$H$_8$)                   & 4.0               & 1.1               \\ 
        $\beta$-Pinene-1S- (C$_{10}$H$_{16}$)   & 3.9               & 0.3               \\  
        DL-Limonene (C$_{10}$H$_{16}$)          & 3.8               & 0.3               \\ 
        Octane (C$_8$H$_{18}$)                  & 3.7               & 0.8               \\ 
        n-Decane (C$_{10}$H$_{22}$)             & 3.6               & 0.7               \\ 
        n-Tridecane (C$_{13}$H$_{28}$)          & 3.6               & 0.6               \\             
        n-Undecane (C$_{11}$H$_{24}$)           & 3.5               & 0.7               \\    
        \Bf{Diethyl sulfide (C$_4$H$_{10}$S)}   & 3.4            & \Bf{2.1}               \\ 
        n-Nonane (C$_9$H$_{20}$)                & 3.3               & 0.8               \\   
        \Bf{DMS (C$_2$H$_6$S)}     & 2.9               & \Bf{2.2}               \\ 
        Diethyl sulfate (C$_4$H$_{10}$O$_4$S)   & 2.7               & 0.8               \\ 
        Dimethyl sulfoxide (C$_2$H$_6$OS)       & 2.4               & 0.4               \\  
        Dimethyl sulfate (C$_2$H$_6$O$_4$S)     & 2.4               & 0.5               \\ 
        Dimethyl disulfide (C$_2$H$_6$S$_2$)    & 2.4               & 1.8               \\ 
        Hexane (C$_6$H$_{14}$)                  & 2.3               & 1.0               \\ 
        \Bf{Chloroethane (C$_2$H$_5$Cl)}        & 2.1               & \Bf{3.2}               \\             
        n-Tetradecane (C$_{14}$H$_{30}$)        & 2.0               & 0.8               \\    \hline
    \end{tabular} 
    \vspace{2mm}
    \caption{The 19 species with $\ln B_{min, L25} \geq 2$ for K2-18~b, and the corresponding $\ln B_{min, M25}$. \Bf{Bold} indicates the only three species that reach $\ln B \geq 2$ across the five spectra from the two works (M25 and L25).
    }
    \label{tab:K2-18b-L25-results}
\end{table}

\subsection{WASP-107~b Results}
\label{subsec:WASP-107b-results}
Finally, we look at WASP-107~b. As mentioned in section~\ref{subsec:exception-WASP-107b}, it has been singled out for its high SNR. The baseline retrieval includes H$_2$O, SO$_2$, H$_2$S, SO and PH$_3$, and we focus on the \texttt{W24} spectrum. We start by computing the evidence for each of these five species by removing them one at a time from the baselines. We show the results in Table~\ref{tab:WASP-107-b-results}. We see that they are preferred with high $\ln B$ values ranging from 7 to 26, so removing just one of them significantly worsens the fit, as shown in Figure~\ref{fig:fit-plot_WASP-107b}. 
We note that SO and PH$_3$ had not been inferred in \cite{Dyrek_Min_2024}, but this work used silicate clouds instead of our simple patchy cloud deck model. Thus, we checked that we were also able to constrain silicate clouds and improve the fit (from $\chi^2 = 76$ to $70$) with this \texttt{pRT} setup, which would likely give us lower evidence values for SO and PH$_3$.
Then, as mentioned in section~\ref{subsec:exception-WASP-107b}, we only look for 44 potential extra absorbers and we find that only six reach $\ln B \geq 0$, and none reach $\ln B \geq 2$. Even if we had looked at all four spectra, we thus know that they would not reach $\ln B_{min} \geq 2$.
There could still be other species reaching the $\ln B \geq 2$, but at least none in common with any other planet.
We note that p-xylene (C$_8$H$_{10}$) and methacrylonitrile come close, at 1.9 and 1.8 respectively. The former was included in the list because we have noticed in previous tests that it can fit the data well, reaching for instance $\ln B = 8.5$ with the same setup applied to the \texttt{CASCADE} spectrum. This is not true for methacrylonitrile which is not preferred, at $\ln B = -0.5$. In Table~\ref{tab:summary-molecules}, we summarize all molecules that are favored at $\ln B_{min} \geq 2$ in at least one planet.

\begin{table}
\renewcommand{\arraystretch}{1.05} 
\setlength{\tabcolsep}{5pt}
\centering
    \begin{tabular}{l|cc} \hline \hline 
        Molecule            & $\ln B$  \\ \hline
        SO$_2$              & \Bf{28.4}     \\    
        H$_2$O              & \Bf{18.6}     \\    
        H$_2$S              & \Bf{17.3}     \\  
        SO                  & \Bf{9.0}      \\
        PH$_3$              & \Bf{7.7}      \\     
        \hline
        p-Xylene            & 1.9      \\        
        Methacrylonitrile   & 1.8      \\    
        CO                  & 0.6      \\
        SiO                 & 0.6      \\ 
        SOF$_2$             & 0.4      \\
        H$_2$CS             & 0.2      \\   \hline 
    \end{tabular} 
    \vspace{2mm}
    \caption{WASP-107~b 5+X results, for the \texttt{W24} spectrum. We first show the evidence for the five baseline species, with \Bf{bold} indicating $\ln B \geq 2$, then the six that reach $\ln B \geq 0$. As mentioned in section~\ref{subsec:WASP-107b-results}, we only look for 44 molecules outside the baseline (see Appendix~\ref{app:species-list}), and none reach $\ln B \geq 2$.
    }
    \label{tab:WASP-107-b-results}
\end{table}

\begin{deluxetable*}{l|cccccccccc}
\renewcommand{\arraystretch}{1.05} 
\setlength{\tabcolsep}{2.2pt}
\tabletypesize{\footnotesize}
\label{tab:summary-molecules}
\tablecaption{Summary of the 19 molecules that reach $\ln B_{min} \geq 2$ in at least one planet. The $\ln B$ values are presented in the same way as in Table~\ref{tab:exploration-uninhabitable}, except for WASP-107~b for which we only use one spectrum, so there is only one $\ln B$ value. We also keep bold font whenever $\ln B_{min} \geq 2$. \newline
$^a$ Indicates that a molecule is included in the baseline for the given planet, so the model comparison is done by removing it from that baseline. We note that the evidence for H$_2$O decreases to $\ln B = 4.0-5.2$ when including SO$_2$ in the model comparison (see Appendix~\ref{app:P24-consitency}).
We see a clear difference between the three temperate sub-Neptunes, where only complex molecules are favored, and the other planets that only show features of simple molecules, if any.
}
\centering
    \tablehead{
        \colhead{Molecule}  & \colhead{K2-18~b} & \colhead{TOI-270~d} & \colhead{TOI-732~c} & \colhead{WASP-39~b} & \colhead{WASP-17~b}& \colhead{WASP-43~b}&\colhead{GJ~1214~b}&\colhead{L~168-9~b}&\colhead{GJ~367~b}&\colhead{WASP-107~b} }
    \startdata 
        Chloroethane        & \Bf{2.1}$-$\Bf{8.0}  & $<0.2$    & \Bf{2.4}$-$\Bf{2.4}&$-$      & $-$     &$<$ 0.2    & $-$     & $-$ & $-$ & $-$ \\ 
        DMS    & \Bf{2.2}$-$\Bf{5.6}  & $<0.4$    & 0.8$-$2.3  &$<$ 0.3   & $-$     &$<$ 0.6    & $-$     & $-$ & $-$ & $-$ \\
        Diethyl sulfide     & \Bf{2.1}$-$\Bf{9.1}  & $<0.1$    & 1.8$-$2.6  &$<$ 0.3   &$-$      &$<$ 0.1    & $-$     & $-$ & $-$ & $-$ \\
        Methacrylonitrile   & -0.8$-$2.9   & \Bf{2.5}$-$\Bf{3.2} &\Bf{2.4}$-$\Bf{4.3}&$<$ 0.2&$-$    &$<$ 0.1    & $-$     & $-$ & $-$ & 1.2 \\  
        Isobutene           & -0.9$-$2.4   & \Bf{2.3}$-$\Bf{4.3} &\Bf{2.3}$-$\Bf{4.8}&$-$    &$-$    &$-$        & $-$     & $-$ & $-$ & $-$ \\  
        Benzonitrile        & $<0.4$       & \Bf{3.4}$-$\Bf{5.3} &$<0.5$      & $<0.5$&$-$    & $<0.9$    & $-$     & $-$ & $0.1$ & $-$\\
        Cyclohexane         & -1.1$-$2.1   & $-$       & \Bf{4.3}$-$\Bf{6.2} &$<$ 0.1 & $-$      &$<$ 1.2    &$-$      & $-$ & $-$ & $-$ \\      
        3-Methylpentane     & -0.9$-$2.2   & $-$       & \Bf{3.7}$-$\Bf{4.3} & $-$    & $-$      & $-$       & $-$     & $-$ & $-$ & $-$ \\            
        2-Butene            & 1.1$-$9.7    & $<1.1$    & \Bf{3.3}$-$\Bf{5.0} &$<$ 0.1 & $-$      &$-$        & $-$     & $-$ & $-$ & $-$ \\    
        3-Methylhexane      & 1.0$-$5.6    & $<0.5$    & \Bf{2.8}$-$\Bf{4.0} &$<$ 0.2 & $-$      & $-$       & $-$     & $-$ & $-$ & $-$ \\          
        1-Pentene           & $<$ 1.5      & $<1.0$    & \Bf{2.4}$-$\Bf{2.5} &$<$ 0.2 & $-$      & $-$       & 0.4  & $-$ & $-$ & $-$ \\      
        cis-2-Pentene       & 0.5$-$4.8    & $<1.3$    & \Bf{2.1}$-$\Bf{2.7} & $-$    & $-$      & $-$       & $-$     & $-$ & $-$ & $-$ \\          
        Hexane              & 1.0$-$6.3    & $<0.1$    & \Bf{2.0}$-$\Bf{5.8} & $-$    & $-$      &$-$        & $-$     & $-$ & $-$ & $-$ \\    
        H$_2$O              & $<0.6$       & $<1.9^a$& $<0.7^a$&\Bf{4.1}$-$\Bf{8.1}$^a$ & 1.8$-$8.0$^a$ & $<0.3^a$ & 0.3 & $-$ & 0.5 & \Bf{18.6}$^a$  \\ 
        SO$_2$              & $-$          & $-$       & $-$  &\Bf{3.1}$-$\Bf{6.3}&$-$       & $-$       & $-$    &$<0.9$ & $-$ &\Bf{28.4}$^a$ \\
        SO$_2$Cl$_2$        & $<0.5$       & $-$       & $-$        & $-$ &\Bf{2.8}$-$\Bf{3.0}&$-$       & $-$     & $-$ & $-$ & $-$ \\ 
        H$_2$S              & $<0.5$       & $-$       & $<0.2$     & $<1.2$      & $-$           & $-$      & $-$  & $-$ & $-$  & \Bf{17.3}$^a$\\
        SO                  & $-$          & $-$       & $-$        & $<1.4$      & $<1.9$        & $-$      & $-$  & $-$ & $-$  & \Bf{9.0}$^a$\\
        PH$_3$              & $<1.3$       & $-$       & $-$        & $<0.1$      & $<0.6$        & $-$      & $-$  & $<0.3$ & $-$  & \Bf{7.7}$^a$\\                  
    \enddata
    \vspace{2mm}
\end{deluxetable*}

\section{Summary and Discussion} 
\label{sec:discussion}
In this work, we carry out a homogeneous survey of 10 JWST MIRI-LRS transmission spectra, each of a different planet. Three of these planets are temperate sub-Neptunes: K2-18~b, TOI-732~c and TOI-270~d, and were predicted to be potential hycean worlds \citep{Madhusudhan_Piette_2021}. The other seven are all hot and/or giant planets, so they cannot be habitable. We compared these observations with the Pearson correlation coefficient, and showed that the temperate sub-Neptunes spectra are very similar to each other, but different from the other spectra. This indicates that the features in their non-flat spectra are unlikely to be due to systematics or random noise.
Then, we performed atmospheric retrievals, and searched for more than 150 potential trace species in each transit observation. When available, we used spectra obtained with different data reductions, in order to improve the robustness of our search, for a total of 21 such reduced spectra explored. Similarly to previous works \citep{Madhusudhan_Constantinou_2025, Rigby_Madhusudhan_2025a, Pica-Ciamarra_Madhusudhan_2026, Welbanks_Nixon_2026, Holmberg_Madhusudhan_2026}, we find that the MIRI spectra of K2-18~b, TOI-732~c and TOI-270~d can each be explained by any one of 2-10 complex molecules (the list of candidates depends on the planet), beyond CH$_4$, CO$_2$ and H$_2$O. 
In contrast, we find no similar evidence for complex molecules in any of the other planets, known to be uninhabitable. Their spectra can be well explained with absorption due to only simple molecules, if any.

\begin{figure*}[t]
    \centering
	\includegraphics[width=0.9\textwidth]{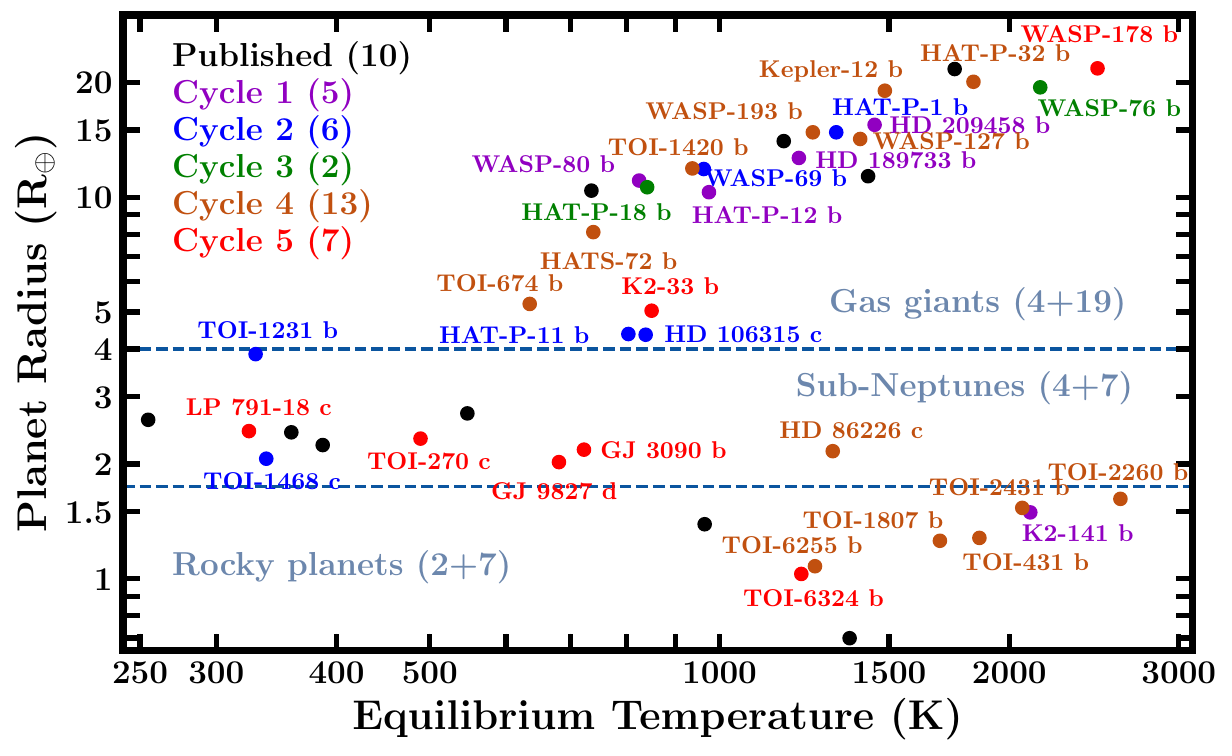}
    \caption{Planets that will be observed with MIRI in transit, in JWST Cycles 1-5. 
    As mentioned in section~\ref{sec:discussion}, we see a wide variety of planet sizes and equilibrium temperatures. 
    The currently published planets are shown in black dots, and listed in Table~\ref{tab:planets-list} and Figure~\ref{fig:planets-plot}.
    We have ommitted the white dwarf planet WD~1856~b (T$_{eq}=160 K$, $R_p = 10.2 R_\oplus$, cycle 4), as well as the disintergrating planets K2-22~b (see section~\ref{subsec:planets}) and BD+05~4868~A~b ($T_{eq} = 1820 K$, unknown R$_p$, cycle 4).
    }
    \label{fig:future-MIRI} 
\end{figure*}

In this survey, we have chosen to focus only on MIRI-LRS spectra. 
There are two main reasons for this choice.
Firstly, it enables us to directly test the exploration approach of previous works \citep{Madhusudhan_Constantinou_2025, Rigby_Madhusudhan_2025a, Pica-Ciamarra_Madhusudhan_2026, Welbanks_Nixon_2026, Holmberg_Madhusudhan_2026}, which were also carried out on MIRI data separately, by applying it to the MIRI spectra of the other seven planets in our sample.
Secondly, a survey of near- and mid-infrared spectra together would not be homogeneous, because not all the planets in the sample have NIR data, and those that do have different numbers of transits with different instruments, as presented in section~\ref{subsec:planets}.
Thus, while combining near- and mid-infrared data is an avenue that should be explored in the future, in order to leverage all the information content available, it is beyond the scope of this present work.
One benefit from this choice is that in this wavelength range ($5-12 \mu$m), the stellar activity is minimal compared to the NIR, which facilitates the identification of planetary spectral features. 
Then, another important limitation of our study is that most of our molecular opacities come from HITRAN cross-section data, which are only available at specific temperature-pressure points (usually T~=~300~K and P~=~1~bar), and air-broadened. More experimental work to extend these at higher temperatures, lower pressures and H$_2$/He broadening would also improve the robustness of our retrievals. 
Finally, as noted in previous works \citep{Rigby_Madhusudhan_2025a, Pica-Ciamarra_Madhusudhan_2026, Holmberg_Madhusudhan_2026}, we have only considered adding single molecules to the baseline species for each planet, so future work could examine different combinations. 

As a consequence of using only MIRI data, which has lower SNR than the NIR data, the current evidence for complex molecules in the atmosphere of temperate sub-Neptunes is low, with values of $\ln B_{min} \approx 2-4$. However, we note that these values are comparable to inferences made on reference targets with similar retrieval architectures and data quality. This includes WASP-39~b, where \cite{Powell2024} inferred molecular absorption from H$_2$O at $\ln B_{min} \approx 2-4$, and SO$_2$ at $\ln B_{min} \approx 3-4$, and we have reproduced consistent values in Appendix~\ref{app:P24-consitency}. This is also similar to the inference of silicate clouds in WASP-17~b \citep{Grant_Lewis_2023}. However, while the silicate clouds were preferred at $\ln B \approx 2$ ($2.6 \sigma$) compared to simple parametric clouds when using the \texttt{ExoTiC} reduction, they were not preferred with the $\ln B$ metric when using the \texttt{Eureka!} spectrum \citep{Grant_Lewis_2023}. This means that in our framework, they would obtain a value of $\ln B_{min} < 2$.

The present results motivate the need for more observations, in order to increase the evidence for both complex molecules in temperate sub-Neptunes as well as simple gas species and/or clouds in other planets.
From our current sample, only TOI-270~d has a new MIRI transit observation planned, with the JWST Cycle 5 GO Program 12157 (PI: B. Benneke). Nevertheless, 35 new planets are being observed in transit with MIRI-LRS in cycles 1 to 5, as shown in Figure~\ref{fig:future-MIRI}. Three of these are temperate sub-Neptunes (TOI-1231~b, TOI-1468~c, and LP~791-18~c), so it will be interesting to see whether they also show hints of extra absorption from complex molecules. The rest of these new planets are very diverse, with 7 rocky planets, 4 warm/hot sub-Neptunes (in addition to the 3 temperate ones) and 19 gas giants. However, we see that there is no temperate gas giant in this sample, which could be interesting to compare with temperate sub-Neptunes. One example of a possible target is TOI-1899~b \citep{Canas_Stefansson_2020}, which has upcoming NIRSpec transit observations (Cycle 2 GO Program 4227, PI: A. Claringbold). There is also no temperate or warm rocky planet, however this is expected as their transmission spectroscopy metrics \citep{Kempton_Bean_2018} are low. Altogether, these upcoming MIRI observations will allow us to extend the comparative characterization of exoplanets as pursued in the present study and establish the robustness of our findings. 

\vspace{3mm} {\it Acknowledgments:} This work is supported by a research grant to N.M. from the UK Research and Innovation (UKRI) Frontier Grant (EP/X025179/1). N.M. and M.B. acknowledge support from the UKRI toward the doctoral studies of M.B (Frontier grant EP/X025179/1).

\vspace{3mm} {\it Author Contributions:} N.M. conceived and led the project. M.B and N.M planned the project. M.B. conducted the spectral comparisons and atmospheric retrievals, and led the writing of the manuscript with inputs from N.M. N.M., M.H. and S.S. provided early access to the reduced data of TOI-270~d from JWST GO Program 3557. All authors provided comments on the manuscript.

\vspace{3mm} {\it Data Availability:} The JWST MIRI-LRS spectra used in this Letter were obtained from previous works, and these sources are listed in Table~\ref{tab:spectra-list}. The raw observational data are published in the Mikulski Archive for Space Telescopes (MAST). Additionnal information including best-fit model spectra and some retrieval outputs are available at https://osf.io/a892w/.

\bibliography{references.bib}
\bibliographystyle{aasjournal}

\appendix

\section{Retrieval Priors}
\label{app:retrieval_priors}
In Table~\ref{tab:retrieval_priors}, we summarize our retrieval priors and baseline species per planet, described in section~\ref{sec:retrieval_setup}. 

\begin{table*}
\centering
\renewcommand{\arraystretch}{1.05} 
\setlength{\tabcolsep}{5.2pt}
\begin{tabular}{lcccccccl} \hline  \hline
    Planet & N$_{free}$ & $\mathrm{log}(P_\mathrm{ref})$ & $T_0$ & $\mathrm{log}(P_\mathrm{c})$ & $\phi$ & $\mathrm{log}(w_{base})$ & {$\mathrm{log}(w_X)$} & Baseline species \\[0.5mm]
    \hline
    WASP-39~b        & 5 & $\mathcal{U}$(-6, 1) & $\mathcal{U}$(500,3000) & $\mathcal{U}$(-6, 1)    & $\mathcal{U}$(0, 1)  & $\mathcal{U}$(-11, 0) & $\mathcal{U}$(-11, 0) & H$_2$O \\
    WASP-17~b        & 5 & $\mathcal{U}$(-6, 1) & $\mathcal{U}$(500,3000) & $\mathcal{U}$(-6, 1)    & $\mathcal{U}$(0, 1)  & $\mathcal{U}$(-11, 0) & $\mathcal{U}$(-11, 0) & H$_2$O \\
    WASP-43~b        & 5 & $\mathcal{U}$(-6, 1) & $\mathcal{U}$(500,3000) & $\mathcal{U}$(-6, 1)    & $\mathcal{U}$(0, 1)  & $\mathcal{U}$(-11, 0) & $\mathcal{U}$(-11, 0) & H$_2$O \\
    GJ~1214~b        & 5 & $\mathcal{U}$(-6, 1) & $\mathcal{U}$(200,1500) & $\mathcal{U}$(-6, 1)    & $\mathcal{U}$(0, 1)  & $\mathcal{U}$(-3, 0)  & $\mathcal{U}$(-11, -1) & CO$_2$  \\
    GJ~367~b         & 5 & $\mathcal{U}$(-6, 1) & $\mathcal{U}$(500,3000) & $\mathcal{U}$(-6, 1)    & $\mathcal{U}$(0, 1)  & $\mathcal{U}$(-3, 0)  & $\mathcal{U}$(-11, -1) & CO$_2$  \\
    L~168-9~b        & 5 & $\mathcal{U}$(-6, 1) & $\mathcal{U}$(200,1500) & $\mathcal{U}$(-6, 1)    & $\mathcal{U}$(0, 1)  & $\mathcal{U}$(-3, 0)  & $\mathcal{U}$(-11, -1) & CO$_2$ \\
    WASP-107~b       & 9 & $\mathcal{U}$(-8, 1) & $\mathcal{U}$(200,1500) & $\mathcal{U}$(-8, 1)    & $\mathcal{U}$(0, 1)  & $\mathcal{U}$(-11, -1) & $\mathcal{U}$(-11, -1) & SO$_2$, H$_2$O, H$_2$S, PH$_3$, SO \\
    K2-18~b          & 6 & $\mathcal{U}$(-6, 1) & $\mathcal{U}$(0,600)    & $\mathcal{U}$(-6, 1)    & $\mathcal{U}$(0, 1)  & $\mathcal{U}$(-11, 0) & $\mathcal{U}$(-11, 0) & CH$_4$, CO$_2$ \\
    TOI-270~d        & 7 & $\mathcal{U}$(-6, 1) & $\mathcal{U}$(0,600)    & $\mathcal{U}$(-6, 1)    & $\mathcal{U}$(0, 1)  & $\mathcal{U}$(-11, 0) & $\mathcal{U}$(-11, 0) & CH$_4$, CO$_2$, H$_2$O \\
    TOI-732~c        & 8 & $\mathcal{U}$(-6, 1) & $\mathcal{U}$(0,600)    & $\mathcal{U}$(-6, 1)    & $\mathcal{U}$(0, 1)  & $\mathcal{U}$(-11, 0) & $\mathcal{U}$(-11, 0) & CH$_4$, CO$_2$, H$_2$O, CO \\ \hline
\end{tabular}

\caption{Priors and baseline species for our retrievals. N$_{free}$ is the number of free parameters in each baseline retrieval, so all other exploration retrievals have one more.
The $R_p$ (to which the radii are fixed) and $M_p$ values are the ones in Table~\ref{tab:planets-list}. See Appendix~\ref{app:sensitivity-bulk-priors} for tested variations to this reference setup.  
}
\label{tab:retrieval_priors}
\end{table*}

\section{Additional Spectra}
\label{app:extra-spectra}
In Table~\ref{tab:extra-spectra}, we list the 4 spectra that were used in the comparisons of section~\ref{sec:observations}, but not in the retrievals (sections~\ref{sec:retrieval_setup}-\ref{sec:results}).
We note that 8 more reductions of the K2-18~b observation were presented in \cite{Stevenson2025}, with different wavelength binning schemes. This means that they cannot be compared as in section~\ref{sec:observations}, and it would take too long to run the retrieval exploration in each of them. Because K2-18~b is already our planet with the most spectra (5), we choose to not study them. 

\begin{table*}
\renewcommand{\arraystretch}{1.05} 
\setlength{\tabcolsep}{7pt}
\centering
\begin{tabular}{llccccccc}
\hline \hline
Planet & Pipeline & N$_p$  & $\lambda$ range & $\chi^2_{flat}$ & $\chi_{r,flat}^2$ & p$_{flat}$ & Ref \\ \hline 
WASP-107~b & \texttt{CASCADE} & 46$^{3:}$ & 5.06-11.83 & 677      & 15.1       & $\leq$$10^{-6}$ & [1] \\  
           & \texttt{Eureka!} & 46$^{3:}$ & 5.06-11.83 & 717      & 15.9       & $\leq$$10^{-6}$ & [1] \\ 
           & \texttt{TEATRO}  & 46$^{3:}$ & 5.06-11.83 & 906      & 20.1       & $\leq$$10^{-6}$ & [1] \\ \hline 
WASP-43~b  & \texttt{Eureka! (v2)} & 14      & 5.25-11.75 & 8.88 & 0.68 & 0.78  & [2] \\  \hline

\end{tabular}
\caption{List of spectra used in the comparisons of section~\ref{sec:observations}, but not in the retrievals (sections~\ref{sec:retrieval_setup}-\ref{sec:results}). 
$^{3:}$ 3 first data points removed (see section~\ref{subsubsec:wavelength-ranges}). 
References: [1] \cite{Dyrek_Min_2024}; [2] \cite{Bell_Crouzet_2024}. 
}
\label{tab:extra-spectra}
\end{table*}

\section{Alternative Metric to Compare Spectra}
\label{app:chi2-comparison}
\begin{figure*}
	\includegraphics[width=0.65\textwidth]{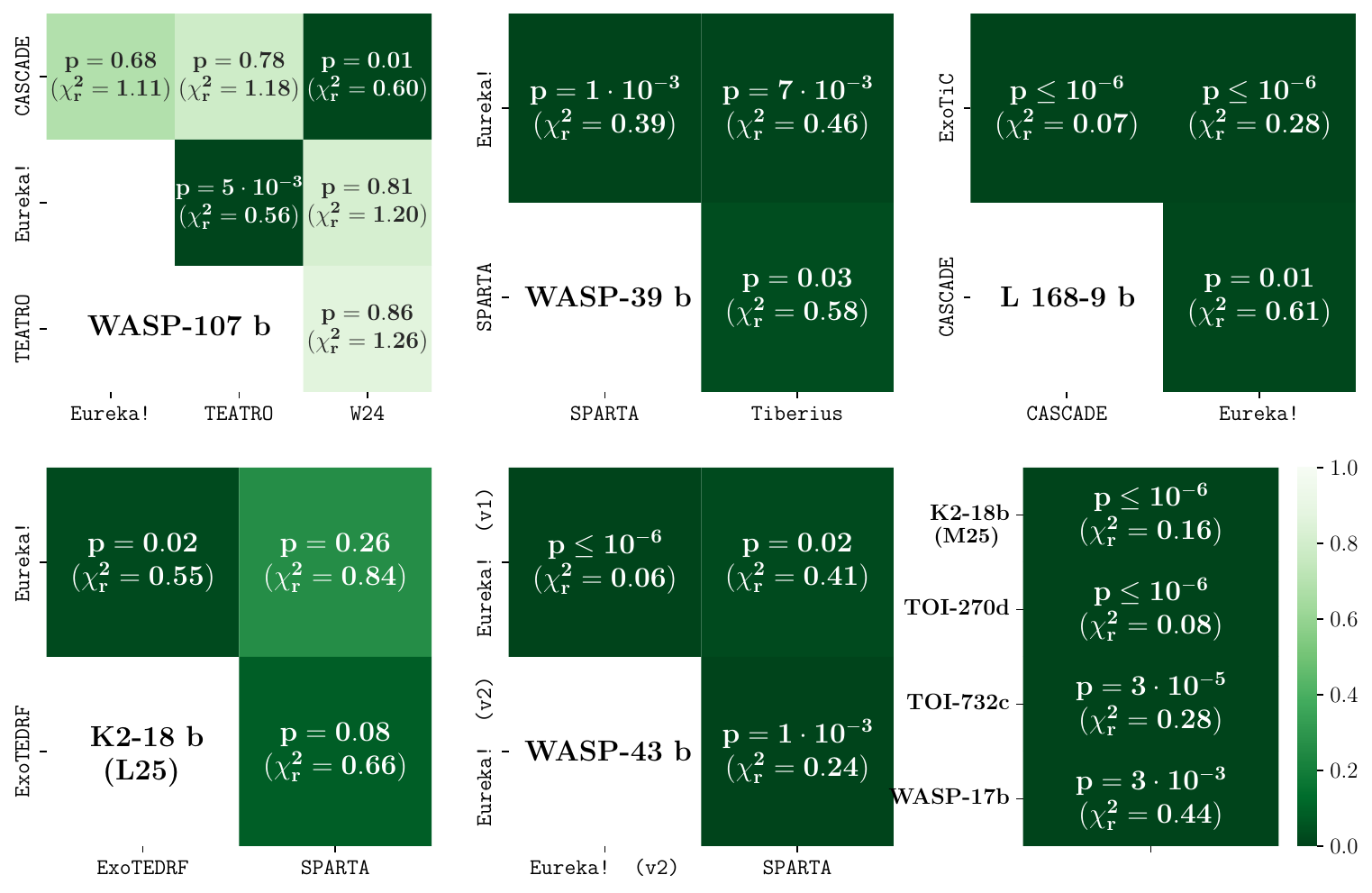}
    \caption{$\chi^2$ and associated p-values between different reductions of the same planet's transit observation.
    }
    \label{fig:correlation-same-planet-chi2diff} 
\end{figure*}

\begin{figure*}
	\includegraphics[width=0.65\textwidth]{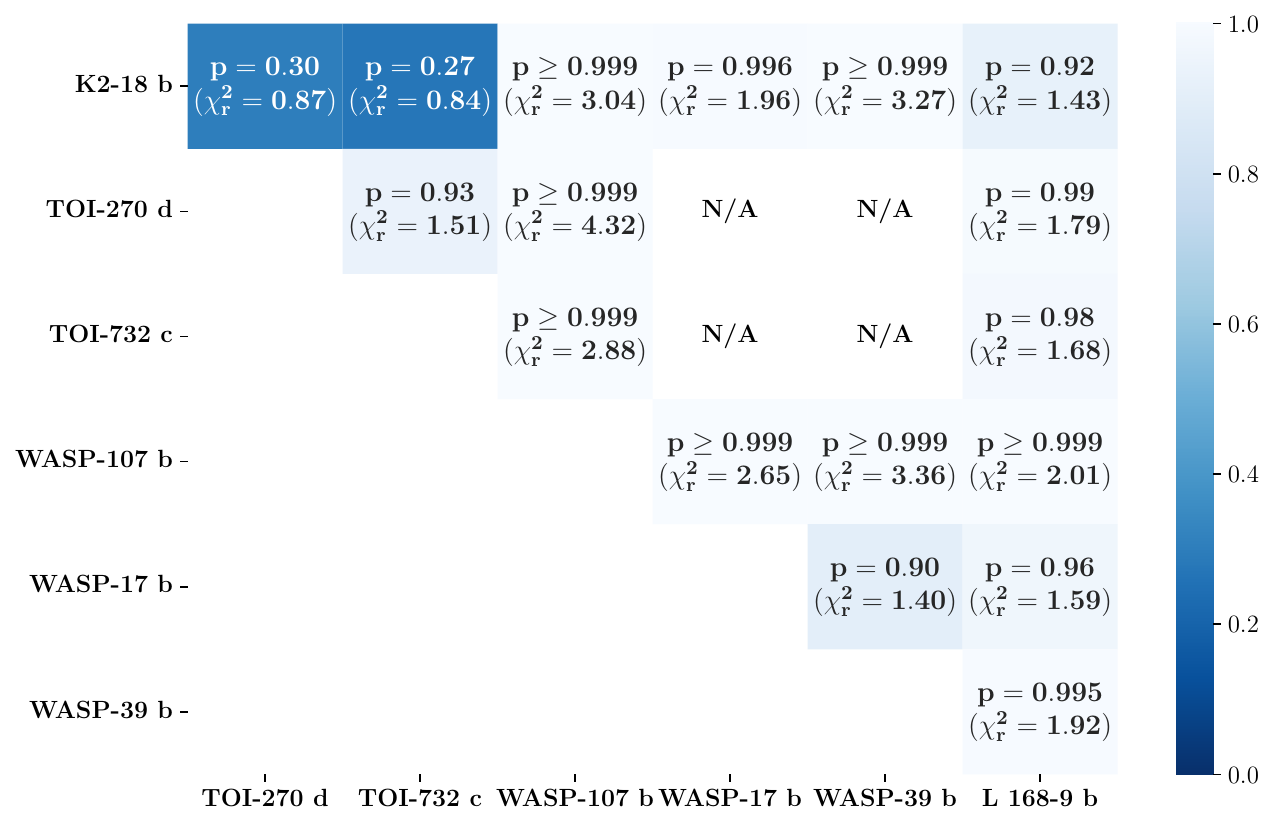}
    \caption{$\chi^2$ and associated p-values between different planets' transit spectra.
    }
    \label{fig:correlation-different-planets-chi2diff} 
\end{figure*}

In sections~\ref{subsubsec:pipeline-consistency}-\ref{subsubsec:planet-correlations}, we have used the Pearson correlation coefficient to compare spectra. Here, we reproduce this study with a different method. We use the spectra converted from transit depths to $\frac{R_p}{R_\star} - (\frac{R_p}{R_\star})_{mean}$, as plotted in Figure~\ref{fig:all-spectra}. For each pair of spectra (X,Y) with errors ($\sigma_X$,$\sigma_Y$), we compute their chi-squared difference $\chi^2_{X,Y} = \frac{(X-Y)^2}{\sigma_X^2 + \sigma_Y^2}$ on their overlapping wavelength grid of $N_p$ bins. From this, we get both $\chi^2_r = \frac{\chi^2_{X,Y}}{N_p - 1}$ and a p-value (this time we don't need to compute it ourselves, we just use the known $\chi^2$ cumulative distribution functions).
With this new metric, we define the p-value as an increasing function of $\chi^2$, so that it is again a decreasing function of the spectra's consistency. This time, two spectra are consistent if p $\lesssim$ 0.95. We show the pipeline comparison results in Figure~\ref{fig:correlation-same-planet-chi2diff}. We see that this metric seems consistent with the Pearson coefficient, as for a given planet the lower $\chi^2_r$ correspond to higher correlations in Figure~\ref{fig:correlation-same-planet}. We also see that every planet has all its p-values below the p = 0.95 threshold (although some pairs of WASP-107~b spectra come close), confirming again that the different reduction pipelines yield consistent data. Finally, in Figure~\ref{fig:correlation-different-planets-chi2diff} we see that K2-18~b is only "strongly consistent" (p $\leq$ 0.5) with the other two temperate sub-Neptunes, and somewhat consistent with L~168-9~b (p = 0.92), as is TOI-270~d with TOI-732~c (p = 0.93). Although it is not as clear as in section~\ref{subsubsec:planet-correlations}, this metric distinguishes again the temperate sub-Neptune spectra from the others.

\section{A Closer Look at WASP-39~\lowercase{b}}
\label{app:WASP-39b}
In this section, we first check the consistency of our WASP-39~b retrievals with those of \cite{Powell2024}, then quantify the degeneracy of SO$_2$ with the other molecules potentially favoured in the planet's atmosphere (see section~\ref{subsubsec:uninhabitable-exploration-result}).
\subsection{Consistency with P24}
\label{app:P24-consitency}
As we said in section~\ref{subsubsec:uninhabitable-planets}, WASP-39~b is the only planet outside of the temperate sub-Neptunes for which retrievals were previously ran on MIRI transit data alone, namely in \cite{Powell2024}. In this previous work, the final retrievals fitted for the planet radius (equivalent to our reference pressure), the isothermal temperature, the log mixing ratios of H$_2$O and SO$_2$, and clouds. The clouds depended on the 7 different retrieval codes used, but 3 of them used the same patchy grey clouds parametrization as our work: \texttt{ARCiS} \citep{Ormel_Min_2019, Min_Ormel_2020}, \texttt{NEMESIS} \citep{Irwin_Teanby_2008, Lee_Fletcher_2012} and \texttt{PyratBay} \citep{Cubillos_Blecic_2021}. So, we can compare the results from these three codes with our own H$_2$O + SO$_2$ retrievals, both in terms of posterior molecular abundances and detection significances. We show this comparison in Table~\ref{tab:WASP-39b-comparison}. We see that all of our median retrieved abundances and most of our $\ln B$ values are within the range found in \cite{Powell2024}. Only the $\ln B$ value for SO$_2$ is lower by 0.6 and 0.2 than their lower bound, for the \texttt{Eureka!} and \texttt{SPARTA} spectra respectively. These differences are lower than the range of their values, so we consider them consistent. This shows that our work is robust to the choice of \texttt{pRT} for the retrievals.

\begin{table}
\renewcommand{\arraystretch}{1.05} 
\setlength{\tabcolsep}{13pt}
\centering
    \begin{tabular}{l|llll} \hline \hline 
        & log(X$_{H_2O}$) & $\ln B$ (H$_2$O) & log(X$_{SO_2}$)& $\ln B$ (SO$_2$) \\ \hline
        & \texttt{Eureka!}   \\ \hline
        This work         & -1.6$_{-0.5}^{+0.3}$ & 5.2 & -5.7$_{-0.5}^{+0.5}$ & 3.4 \\
        P24               & -1.5 to -1.6         & 2.0 to 9.8  & -5.4 to -5.8  & 4.0 to 4.8 \\ \hline
        & \texttt{Tiberius}  \\ \hline
        This work         & -2.0$_{-0.7}^{+0.4}$ & 4.0 & -5.6$_{-0.6}^{+0.5}$ & 6.3 \\      
        P24               & -1.9 to -2.1         & 2.2 to 9.3  & -5.6 to -5.7         & 5.5 to 7\\  \hline
        & \texttt{SPARTA}    \\ \hline
        This work         & -1.5$_{-0.7}^{+0.3}$ & 4.3 & -5.5$_{-0.6}^{+0.5}$ & 3.1 \\       
        P24               & -1.3 to -1.7         & 2.1 to 4.6  & -5.3 to -5.6         & 3.3 to 4.0 \\ \hline
    \end{tabular} 
    \vspace{2mm}
    \caption{WASP-39~b comparison between our work and the \texttt{ARCiS}, \texttt{NEMESIS} and \texttt{PyratBay} retrievals of \cite{Powell2024}. In this previous work, the model preferences were given as $\sigma$ values, so we convert them to $\ln B$ values using the convention from \cite{Sellke_Bayarri_2001} (see Appendix~\ref{app:sigma-conversion}).
    }
    \label{tab:WASP-39b-comparison}
\end{table}

\subsection{WASP-39~b: Degeneracy with SO\textsubscript{2}} 
\label{app:WASP-39b-SO2}

\begin{table}
\renewcommand{\arraystretch}{1.05} 
\setlength{\tabcolsep}{6pt}
\centering
    \begin{tabular}{l|ll} \hline \hline 
        Molecule      & 1+X     & 2+X  \\ \hline
        SO$_2$        & 6.3     & N/A  \\   
        CH$_4$        & 5.3     & 0.5  \\ 
        SO$_3$        & 3.4     & 0.3  \\ 
        SOF$_2$       & 2.9     & -0.1 \\  
        C$_2$H$_2$    & 2.2     & 0.4  \\ 
        CH$_3$Cl      & 2.1     & 0.5  \\ 
        HCN           & 2.0     & 0.3  \\ 
        Tiophene      & 2.0$^a$ & 0.9  \\             
        Cumene        & 1.8     & 0.1  \\    
        SO            & 1.4     & 0.5  \\ 
        H$_2$S        & 1.2     & 0.2  \\    
        CS            & 1.1     & 0.3  \\ \hline
    \end{tabular} 
    \vspace{2mm}
    \caption{Maximum evidence $\ln B_{max}$ values obtained for the WASP-39~b 2+X retrieval results, as described in Appendix~\ref{app:WASP-39b-SO2}, compared to the 1+X results of section~\ref{subsubsec:uninhabitable-exploration-result}.
    $^a$ Tiophene (C$_4$H$_4$S) was not listed in Table~\ref{tab:exploration-uninhabitable} because it has $\ln B_{max} = 1.97 \leq 2$.
    }
    \label{tab:WASP-39b-2+X}
\end{table}

As mentioned previously, SO$_2$ was robustly identified in WASP-39~b \cite{Powell2024} with all three data reductions. In this work, we find that its \texttt{Tiberius} spectrum can also be fitted at $\ln B \geq 2$ in the H$_2$O+X setup by 6 other molecules X. These include CH$_4$ at $\ln B = 5.3$, which is close to the $\ln B = 6.3$ reached by SO$_2$. However, we see in Figure~\ref{fig:fit-plot_WASP-39b} that the preference for CH$_4$ is mostly due to the data points at $7.5-8 \mu m$, and this is also true for the other species. So, they could be degenerate with SO$_2$. To evaluate this, we use a new retrieval setup with both SO$_2$, one of these species (still named X), and we keep H$_2$O. We refer to this setup as 2+X, in contrast to the previous 1+X (or H$_2$O+X) setup. We run this setup on all three spectra, for the 6 molecules that reached $\ln B_{max} \geq 2$ in the 1+X setup, as well as the 5 others with $\ln B_{max} \geq 1$. We show the results in Table~\ref{tab:WASP-39b-2+X}.

We see that none of these 11 species is preferred at $\ln B \geq 1$ for any spectrum when SO$_2$ is included in the baseline. The highest preference with this 2+X setup is $\ln B = 0.9$, reached for tiophene (C$_4$H$_4$S) with the \texttt{Tiberius} spectrum. Even for this retrieval, the VMR constraints on H$_2$O and SO$_2$ are changed by less than 0.2 dex compared to the H$_2$O+SO$_2$ retrieval (as well as the other posteriors), whereas tiophene is highly unconstrained at a -6.3$_{-3.9}^{+1.0}$. This analysis shows that although other species could potentially fit the WASP-39~b MIRI data, their evidence disappears when adding SO$_2$. As mentioned in section~\ref{subsubsec:uninhabitable-planets}, SO$_2$ was further confirmed by the NIR data (see section~\ref{subsubsec:uninhabitable-planets}), and none of these other species was found (apart from potentially H$_2$S), including CH$_4$ which has strong features in the NIR. In summary, SO$_2$ is by far the best explanation for the $7.5-8 \mu m$ feature seen in the MIRI transit of WASP-39~b, and no extra molecule is needed.

\section{Conversion from B to $\sigma$-Values}
\label{app:sigma-conversion}
As mentioned in section~\ref{sec:results}, we chose to not convert our $\ln B$ evidences into $\sigma$ detection significances, following the recommendation of \cite{Thorngren_Sing_2026}. For completeness, we show a reminder of this conversion in Table~\ref{tab:sigma-conversion}.

\begin{table}
\renewcommand{\arraystretch}{1.05} 
\setlength{\tabcolsep}{13pt}
\centering
    \begin{tabular}{ccl} \hline \hline 
        $\ln B$   & N$\sigma$   & Strength of evidence  \\ \hline
        0         & 1.0         & None                  \\
        1         & 2.0         & Weak                  \\
        2         & 2.5         & Weak to moderate      \\
        3         & 3.0         & Moderate              \\
        5         & 3.6         & Strong                \\
        10        & 4.9         & Definitive            \\ \hline
    \end{tabular} 
    \vspace{2mm}
    \caption{Conversion between $\ln B$ and $\sigma$ values from \cite{Sellke_Bayarri_2001}, and strength of evidence from \cite{Trotta2008}. 
    }
    \label{tab:sigma-conversion}
\end{table}

\section{BPICS Metric for Model Preference}
\label{app:BPICS}
In this section, we compute the model preference for our retrievals using an alternative metric to the Bayes factor. We use the BPICS, introduced in \cite{Ando_2011} and recommended in \cite{Thorngren_Sing_2026}:
$$BPICS = -2E_\theta[\ln(p(y|\theta)] + 2k.$$
Here, $\ln(p(y|\theta)$ is the likelihood of the data y with the parameter values $\theta$, $E_\theta$ is the mean across posterior samples, and k is the number of free parameters. When comparing two retrieval models M$_1$ and M$_2$, one computes $\Delta BPICS = BPICS_{M2} - BPICS_{M1}$, and $M2$ is preferred if $\Delta BPICS \leq 0$. For an easier comparison with the $\ln B$ metric, in what follows we will actually provide values of $-\frac{1}{2} BPICS = E_\theta[\ln(p(y|\theta)] - k$, which we will refer to as $\overline{BPICS}$. 

Thus, the model preference for a given molecule X is defined the same way as $\ln B$: if there are N baseline molecules, it is $\Delta \overline{BPICS}$ = $\overline{BPICS}_{N+X}$ - $\overline{BPICS}_N$.
We start with a side-to-side comparison of $\ln B$ and $\Delta \overline{BPICS}$ values for five spectra of K2-18~b in Table~\ref{tab:K2-18b-BPICS}, for the 12 molecules that have $\ln B_{min} \geq 1$. First, we see that both metrics scale approximately the same way, which confirms our choice of multiplying the BPICS values by $-\frac{1}{2}$.
However, for an equal threshold, the BPICS is slightly more conservative: only 5 molecules have $\Delta \overline{BPICS}_{min} \geq1$, and DMS has $\Delta \overline{BPICS}_{min} \geq2$, diethyl sulfide and chloroethane now falling below the threshold.

Then, we compute this metric on all the other planets' retrievals, and find again mostly consistent values between $\ln B$ and $\overline{BPICS}$, summarized in Table~\ref{tab:summary-BPICS}. However, just like in K2-18~b, chloroethane has $\Delta \overline{BPICS_{min}} \leq 2$ in TOI-732~c, and only benzonitrile remains above the threshold in TOI-270~d. On the other hand, $H_2O$ and 2,4,4-trimethyl-1-pentene are now above the threshold in WASP-17~b and TOI-732~c respectively. Finally, methacrylonitrile and p-xylene now have $\Delta \overline{BPICS} \geq 2$ in the \texttt{W24} spectrum of WASP-107~b. But just like we found in section~\ref{subsec:WASP-107b-results}, only p-xylene is also favored in the \texttt{CASCADE} spectrum.

\begin{table}
\renewcommand{\arraystretch}{1.05} 
\setlength{\tabcolsep}{11pt}
\centering
    \begin{tabular}{l|cccccc} \hline \hline 
        Molecule & Minimum & \texttt{JExoRES} & \texttt{JexoPipe} & \texttt{Eureka!} & \texttt{ExoTEDRF} & \texttt{SPARTA} \\ \hline
        ln Z$_{base}$ & N/A & 213.08 & 207.97 & 193.48 & 195.94 & 194.45  \\ 
        $\overline{BPICS}_{base}$ & N/A & 208.46 & 203.45 & 189.49 & 192.15 & 192.40 \\
        \hline
        \und{\bf{Dimethyl sulfide}}& \und{\Bf{2.2 / 2.1}}& 2.2 / 2.1 & 2.8 / 3.4 & 5.6 / 6.8 & 2.9 / 3.1 & 4.6 / 4.4 \\          
        \bf{DMS}   & \Bf{2.1} / 1.9& 2.1 / 1.9 & 2.7 / 3.2 & 9.1 / 11.0& 5.5 / 6.1 & 3.4 / 2.9             \\           
        \bf{Chloroethane}      & \Bf{2.1} / 1.3& 3.2 / 3.5 & 4.4 / 5.4 & 8.0 / 9.2 & 4.7 / 5.2 & 2.1 / 1.3             \\      
        Butane                 & 1.9 / 1.4     & 2.6 / 2.6 & 2.6 / 2.9 & 9.2 / 10.2& 4.5 / 5.0 & 1.9 / 1.4             \\ 
        Dimethyl disulfide     & 1.8 / 1.5     & 1.8 / 1.5 & 2.1 / 2.1 & 5.5 / 6.3 & 2.4 / 2.2 & 5.5 / 5.1             \\   
        1-Chloropentane        & 1.6 / 0.8     & 1.6 / 0.9 & 2.4 / 2.7 & 4.2 / 4.6 & 1.7 / 1.4 & 1.8 / 0.8             \\            
        Ethyl mercaptan        & 1.6 / 1.0     & 1.6 / 1.0 & 2.7 / 3.2 & 7.9 / 9.2 & 6.3 / 7.3 & 6.1 / 6.4             \\        
        Dichloromethylphosphine& 1.2 / 0.2     & 1.4 / 0.7 & 2.1 / 2.0 & 1.7 / 1.8 & 1.2 / 0.6 & 1.6 / 0.2             \\ 
        Heptane                & 1.1 / 0.2     & 1.1 / 0.2 & 1.2 / 0.6 & 5.7 / 7.0 & 2.3 / 2.4 & 1.2 / 1.2             \\    
        Bromoethane            & 1.1 / 0.1     & 2.5 / 2.5 & 2.0 / 2.1 & 8.8 / 10.3& 6.2 / 7.0 & 1.1 / 0.1             \\       
        2-Butene               & 1.1 / 0.6     & 2.2 / 2.3 & 1.1 / 0.6 & 9.7 / 11.9& 4.4 / 5.1 & 4.0 / 3.8             \\  
        Hexane                 & 1.0 / 0.1     & 1.0 / 0.1 & 1.3 / 0.8 & 6.3 / 7.4 & 2.3 / 2.2 & 2.5 / 2.8             \\   \hline
    \end{tabular} 
    \vspace{2mm}
    \caption{K2-18~b model preference per molecule and spectrum: $\ln B /\Delta \overline{BPICS}$} \Bf{Bold} indicates the 3 molecules with $\ln B_{min} \geq 2$. \und{Underlined} indicates DMS, the only molecule with $\Delta \overline{BPICS}_{min} \geq 2$.
    \label{tab:K2-18b-BPICS}
\end{table}

\section{Sensitivity to Bulk Parameter Priors}
\label{app:sensitivity-bulk-priors}
In this section, we test the sensitivity of our retrieval results to the selected priors, as described in sections \ref{subsec:temp-and-clouds} and \ref{subsec:bulk-parameters}.
The three changes we test are adding a Gaussian prior on the planet mass, with the uncertainties reported in Table~\ref{tab:planets-list}; replacing the reference pressure prior by a uniform prior prior on the planet radius, between 0.8 and 1.2 $R_p$; and narrowing the temperature priors.
For this, we pick four different combinations of planet and extra absorber that are of particular interest: DMS on K2-18~b, isobutene on TOI-270~d, SO$_2$ on WASP-39~b and WASP-107~b. The interest for the latter is because it is the highest model preference found in our work ($\ln B = 28.4$). For K2-18~b and TOI-270~d, we pick the \texttt{JExoRES} reduction, as it is the one that causes the lowest $\ln B$ values for DMS and isobutene respectively. We present our results in Table~\ref{tab:sensitivity-bulk}. We see that the changes in the baseline $\ln Z$ and $\overline{BPICS}$ values are small ($\lesssim2$), and more importantly that the changes in $\ln B$ and $\Delta \overline{BPICS}$ for each molecule are even smaller ($\lesssim0.5$, except for WASP-107~b in which case they are $\lesssim1$). In particular, DMS and isobutene both remain at $\ln B \geq 2$ regardless of the retrieval prior choices, and DMS also at $\Delta\overline{BPICS} \geq 2$ (as observed in Appendix~\ref{app:BPICS}, isobutene does not reach that threshold). We conclude that most of the inferences made in this work are unlikely to be affected by these different prior choices.

\section{Spectral Fits}
\label{app:spectral-fits}
\begin{figure*}
\centering
	\includegraphics[width=0.61\textwidth]{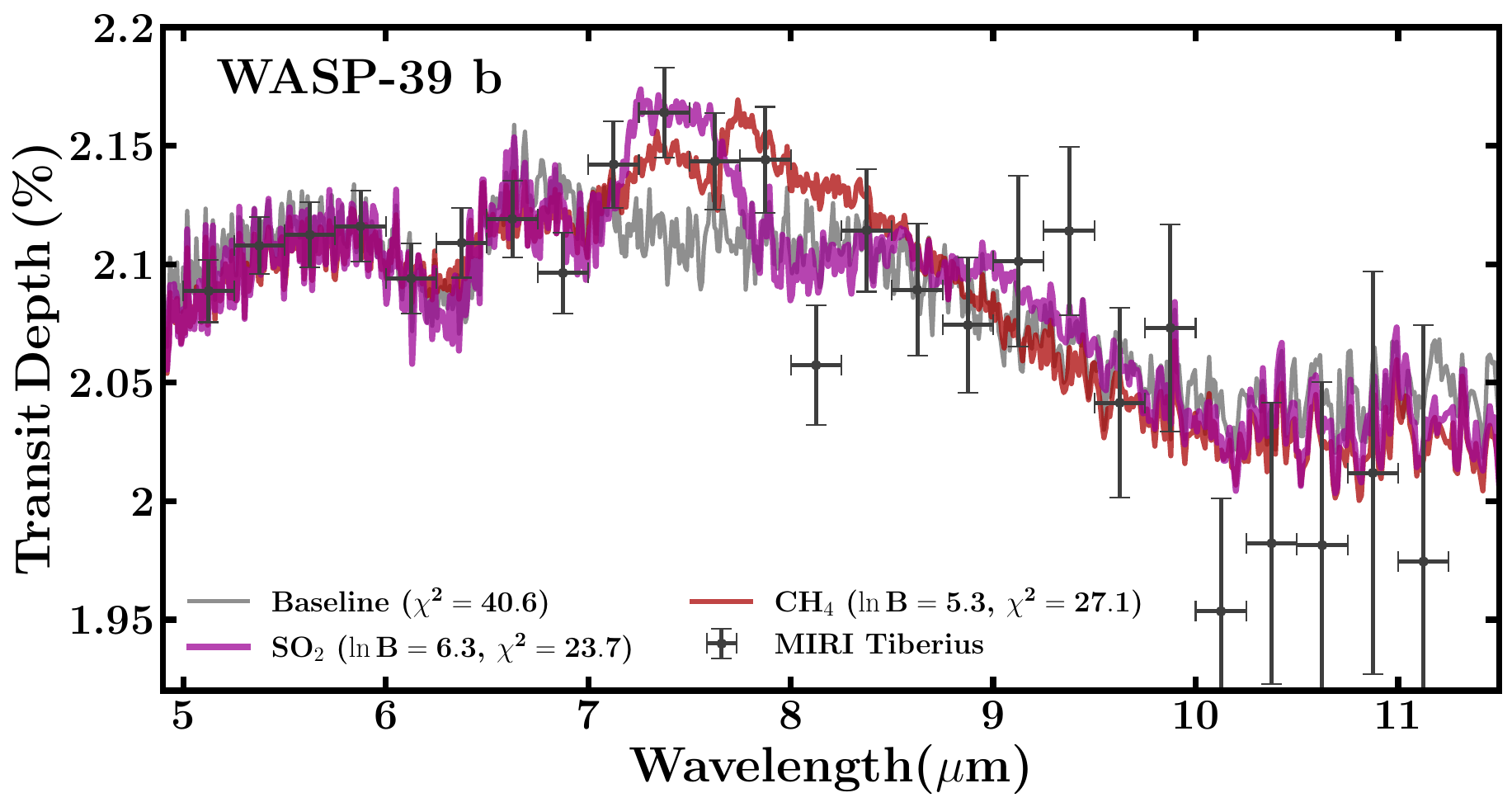}
     \caption{
     Retrieved spectra with the WASP-39~b \texttt{Tiberius} data. The solid curves show best-fit retrieved spectra obtained from the \texttt{pRT} baseline retrieval (including H$_2$O), and the ones with added SO$_2$ and CH$_4$, as described in section~\ref{subsubsec:uninhabitable-exploration-result}. We see that SO$_2$ and CH$_4$ both improve the fit in a similar wavelength range ($\approx 7.5-8 \mu m$), hence their partial degeneracy described in Appendix~\ref{app:WASP-39b-SO2}.
     We note that the last data point at 11.375 $\mu m$ is below the figure, with a transit depth of $1.75_{-0.117}^{+0.117}\%$.
     }
     \label{fig:fit-plot_WASP-39b} 
\end{figure*}

\begin{figure*}
\centering
	\includegraphics[width=0.61\textwidth]{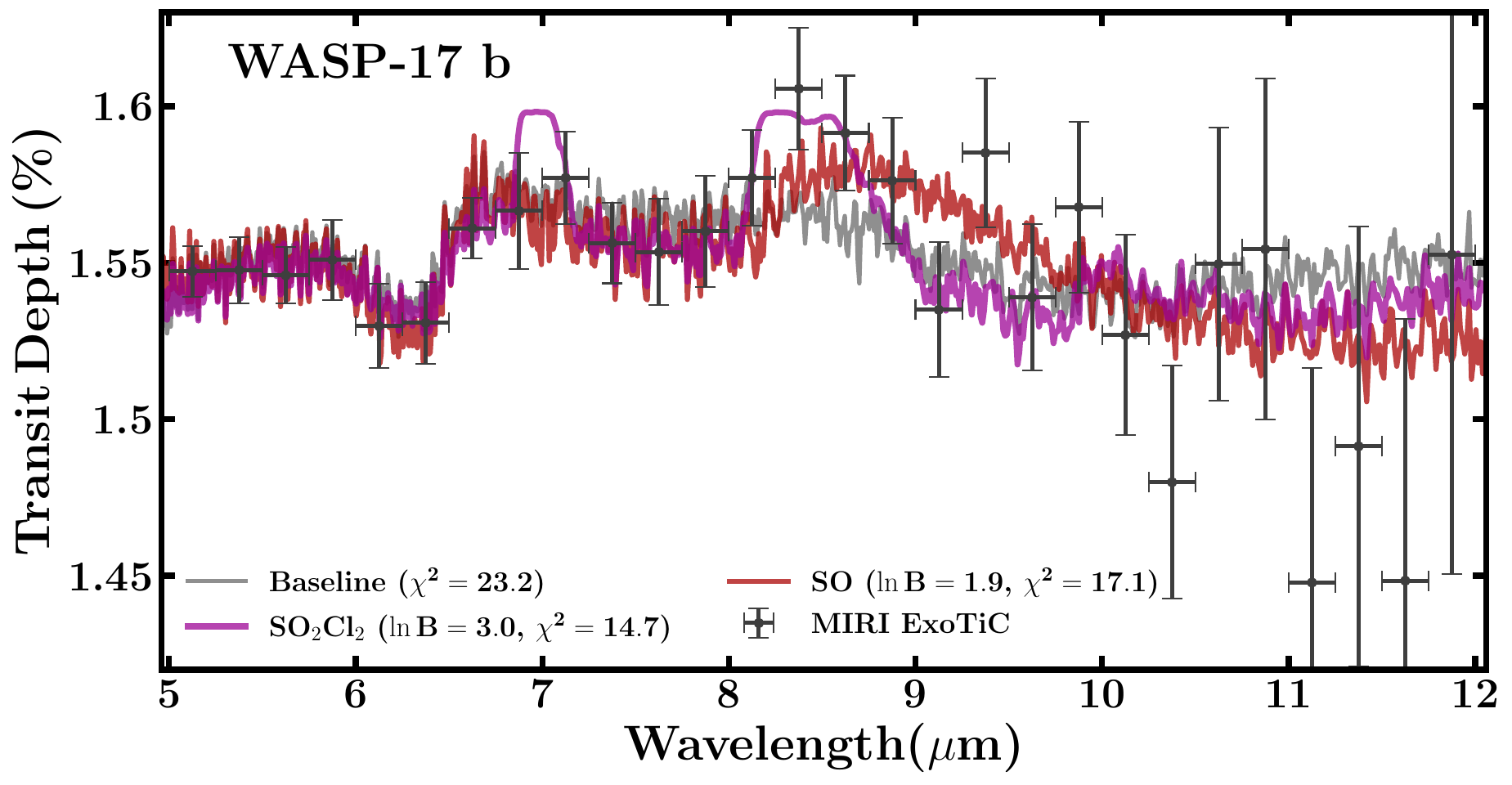}
     \caption{
     Retrieved spectra with the WASP-17~b \texttt{ExoTiC} data. The solid curves show best-fit retrieved spectra obtained from the \texttt{pRT} baseline retrieval (including H$_2$O), and the ones with added SO$_2$Cl$_2$ and SO, as described in section~\ref{subsubsec:uninhabitable-exploration-result}. 
     }
     \label{fig:fit-plot_WASP-17b} 
\end{figure*}

\begin{figure*}
\centering
	\includegraphics[width=0.61\textwidth]{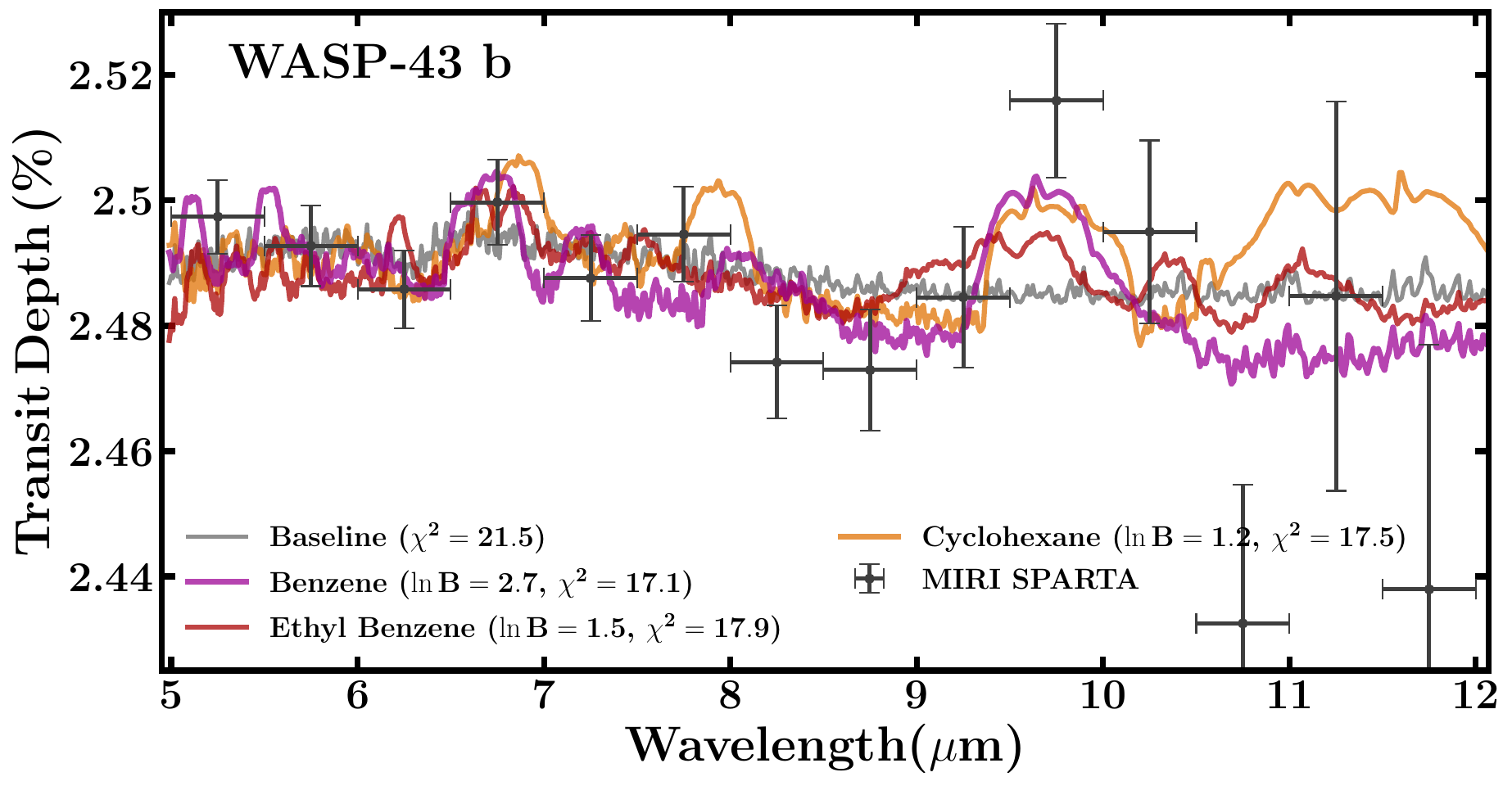}
     \caption{
     Retrieved spectra with the WASP-43~b \texttt{SPARTA} data. The solid curves show best-fit retrieved spectra obtained from the \texttt{pRT} baseline retrieval (including H$_2$O), and the ones with added benzene, ethyl benzene and cyclohexane, as described in section~\ref{subsubsec:uninhabitable-exploration-result}. 
     }
     \label{fig:fit-plot_WASP-43b} 
\end{figure*}

\begin{figure*}
\centering
	\includegraphics[width=0.61\textwidth]{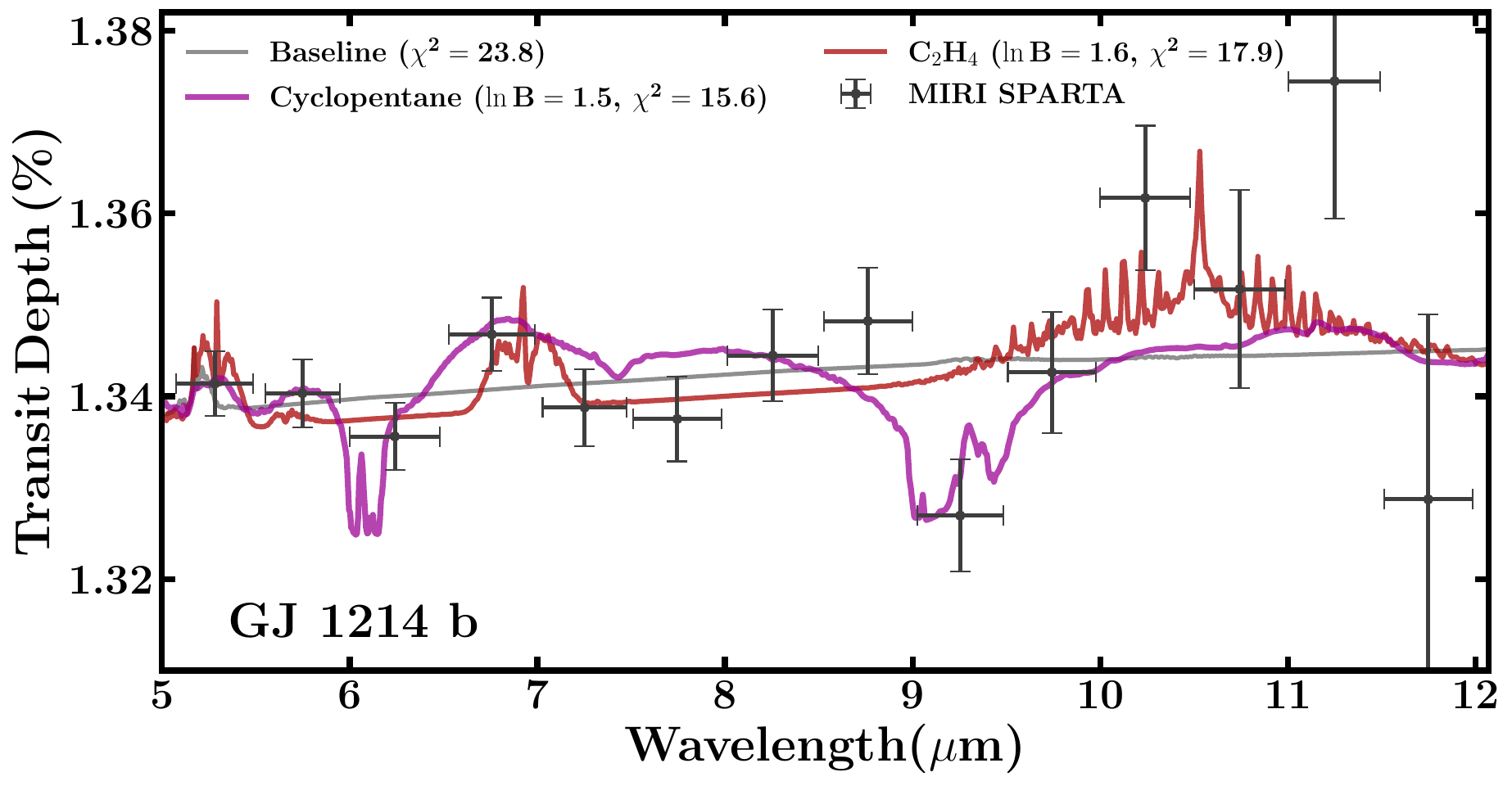}
     \caption{
     Retrieved spectra with the GJ~1214~b \texttt{SPARTA} data. The solid curves show best-fit retrieved spectra obtained from the \texttt{pRT} baseline retrieval (including CO$_2$), and the ones with added cyclopentane and C$_2$H$_4$, as described in section~\ref{subsubsec:uninhabitable-exploration-result}.
     We note that previous studies excluded the last 4 data points because of potential correlated noise, but we chose to keep them.
     }
     \label{fig:fit-plot_GJ1214b} 
\end{figure*}

\begin{figure*}
\centering
	\includegraphics[width=0.61\textwidth]{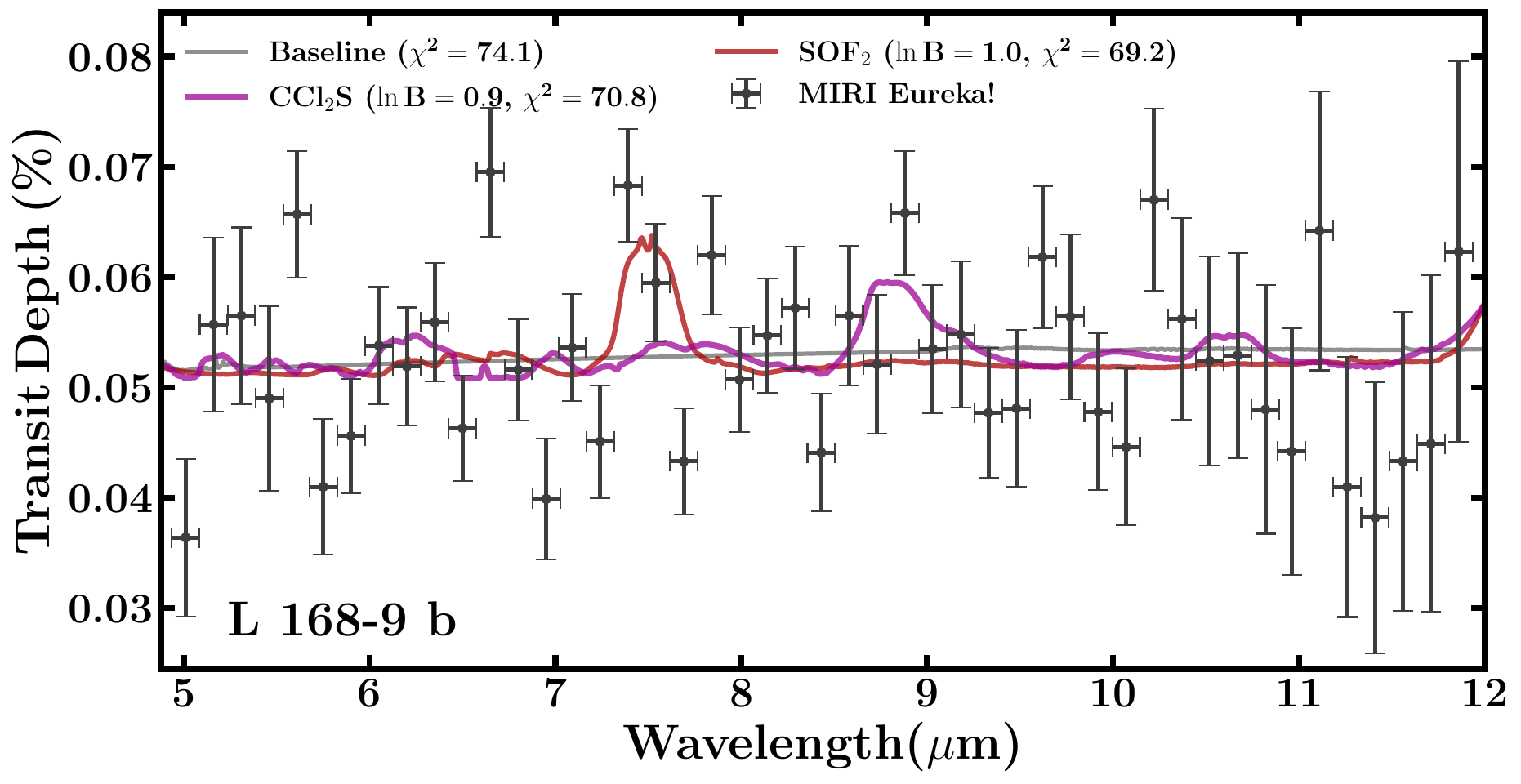}
     \caption{
     Retrieved spectra with the L~168-9~b \texttt{Eureka!} data. The solid curves show best-fit retrieved spectra obtained from the \texttt{pRT} baseline retrieval (including CO$_2$), and the ones with added CCl$_2$S and SOF$_2$, as described in section~\ref{subsubsec:uninhabitable-exploration-result}.
     }
     \label{fig:fit-plot_L168-9b} 
\end{figure*}

\begin{figure*}
\centering
	\includegraphics[width=0.61\textwidth]{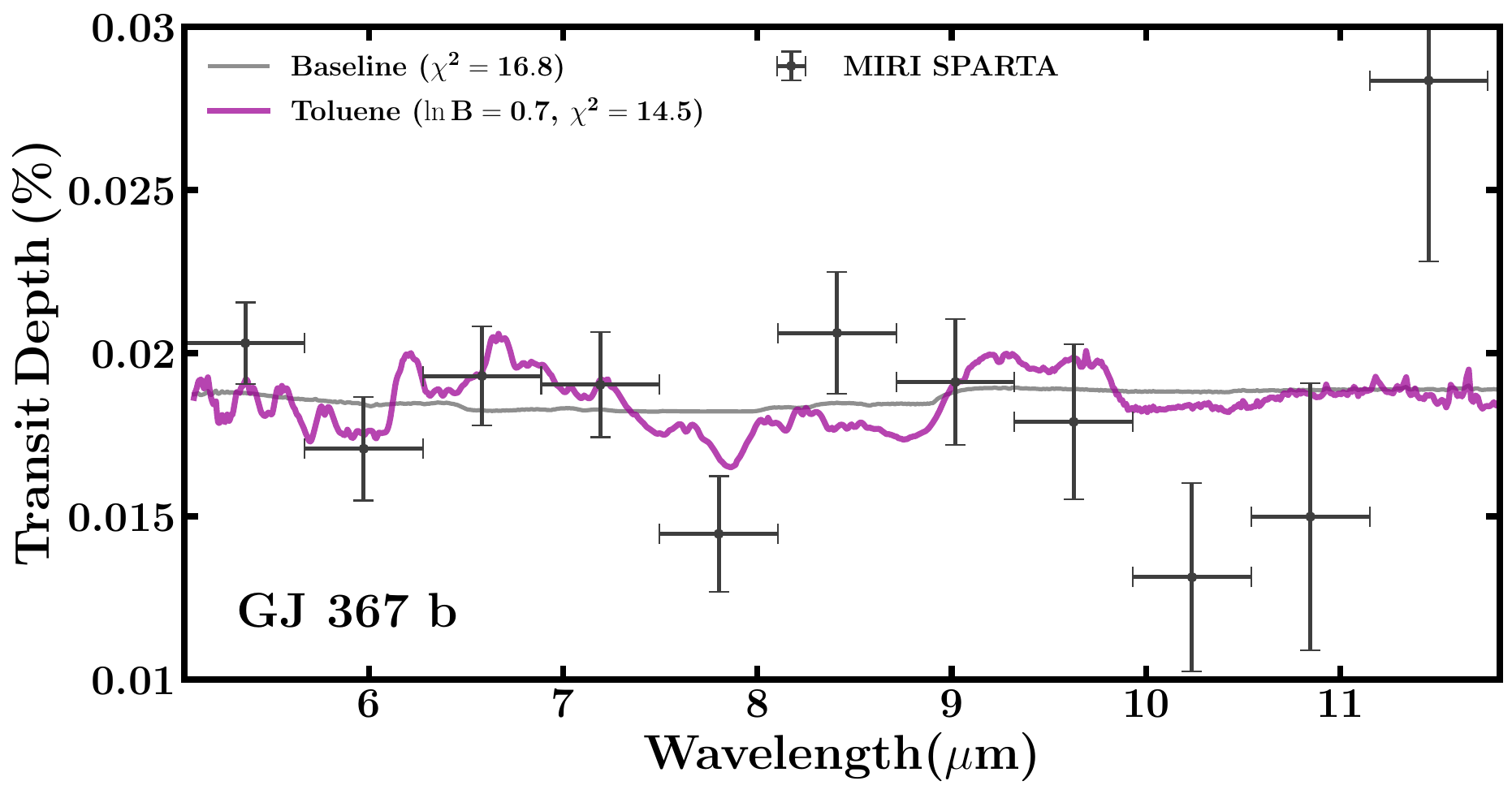}
     \caption{
     Retrieved spectra with the GJ~367~b \texttt{SPARTA} data. The solid curves show best-fit retrieved spectra obtained from the \texttt{pRT} baseline retrieval (including CO$_2$), and the one with added toluene, as described in section~\ref{subsubsec:uninhabitable-exploration-result}.
     }
     \label{fig:fit-plot_GJ-367-b} 
\end{figure*}

\begin{figure*}
\centering
	\includegraphics[width=0.61\textwidth]{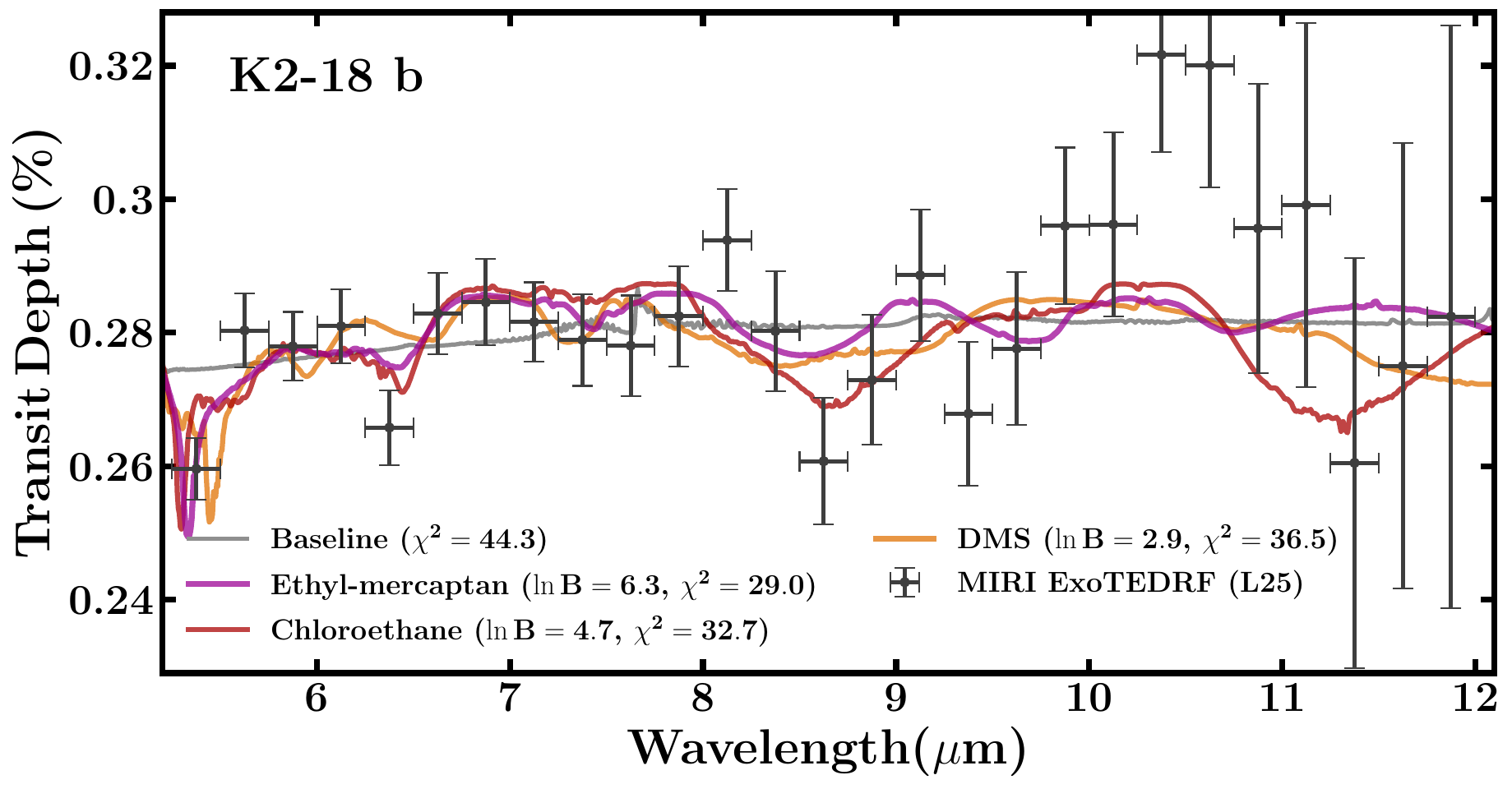}
     \caption{
     Retrieved spectra with the K2-18~b \texttt{ExoTEDRF} data. The solid curves show best-fit retrieved spectra obtained from the \texttt{pRT} baseline retrieval (including CH$_4$ and CO$_2$), and the ones with added ethyl mercaptan, chloroethane and DMS, as described in section~\ref{subsubsec:previous-sub-Neptunes}.
     }
     \label{fig:fit-plot_K2-18b} 
\end{figure*}

\begin{figure*}
\centering
	\includegraphics[width=0.61\textwidth]{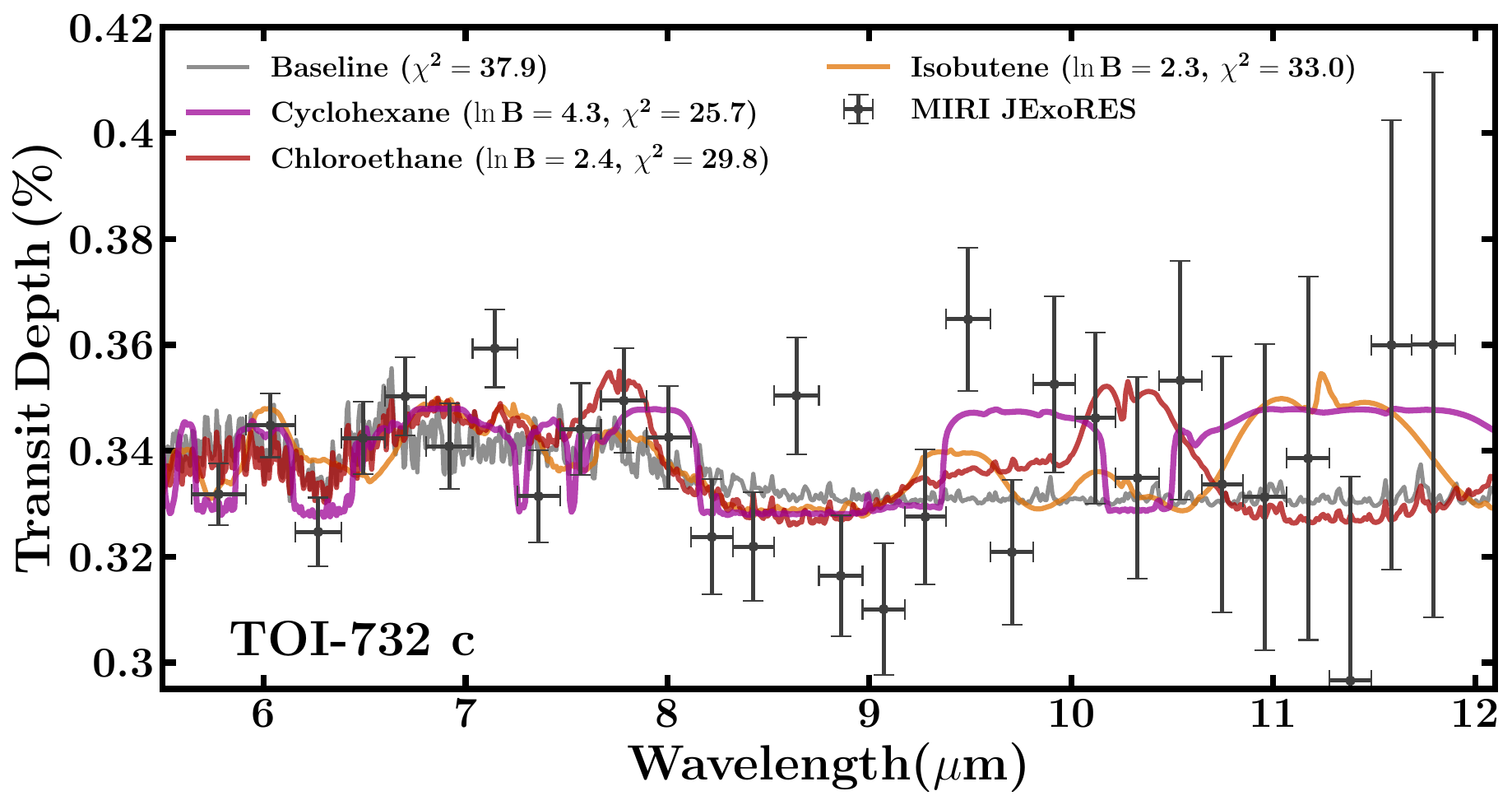}
     \caption{
     Retrieved spectra with the TOI-732~c \texttt{JExoRES} data. The solid curves show best-fit retrieved spectra obtained from the \texttt{pRT} baseline retrieval (including CH$_4$, CO$_2$, H$_2$O and CO), and the ones with added cyclohexane, chloroethane and isobutene, as described in section~\ref{subsubsec:previous-sub-Neptunes}.
     }
     \label{fig:fit-plot_TOI-732c} 
\end{figure*}

\begin{figure*}
\centering
	\includegraphics[width=0.61\textwidth]{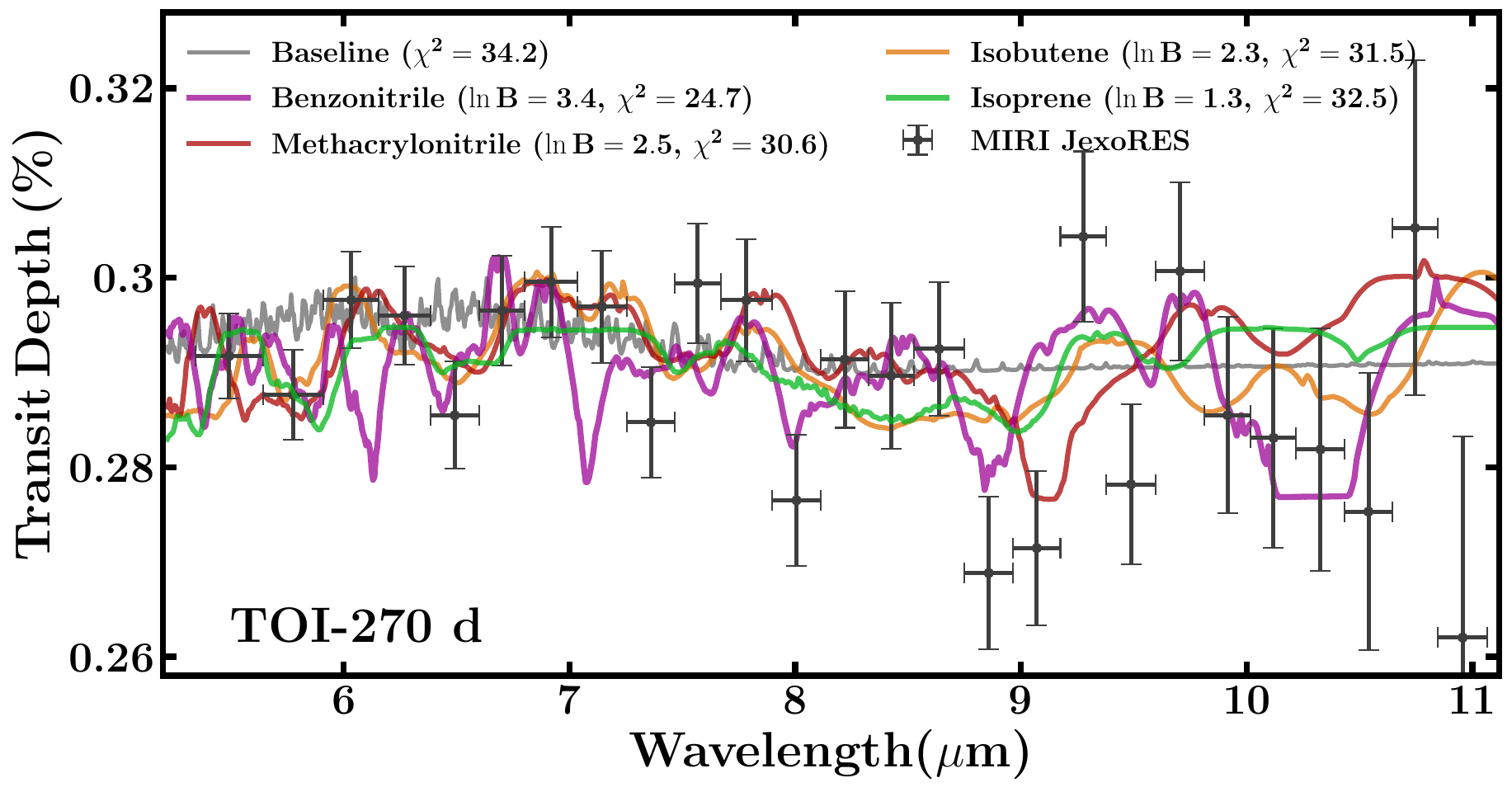}
     \caption{
     Retrieved spectra with the TOI-270~d \texttt{JExoRES} data, (same as Figure 4 of \citealt{Holmberg_Madhusudhan_2026}). The solid curves show best-fit retrieved spectra obtained from the \texttt{pRT} baseline retrieval (including CH$_4$, CO$_2$ and H$_2$O), and the ones with added benzonitrile, methacrylonitrile, isobutene and isoprene, as described in section~\ref{subsubsec:previous-sub-Neptunes}.
     }
     \label{fig:fit-plot_TOI-270d} 
\end{figure*}

\begin{figure*}
\centering
	\includegraphics[width=0.61\textwidth]{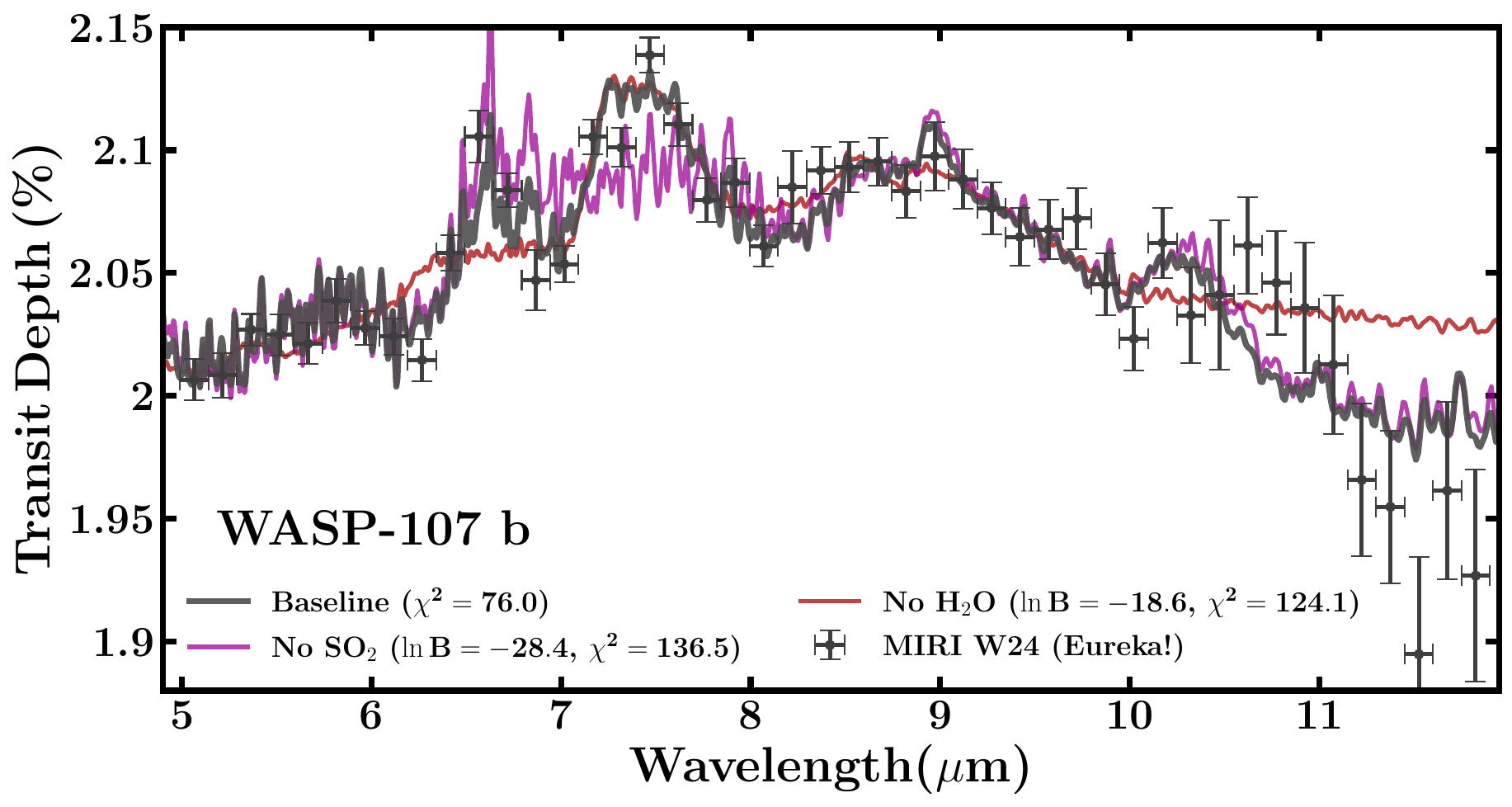}
     \caption{
     Retrieved spectra with the WASP-107~b \texttt{W24} data. The solid curves show best-fit retrieved spectra obtained from the \texttt{pRT} baseline retrieval (including SO$_2$, H$_2$O, H$_2$S, PH$_3$ and SO), and the ones with removed SO$_2$ and H$_2$O, as described in section~\ref{subsec:WASP-107b-results}.
     }
     \label{fig:fit-plot_WASP-107b} 
\end{figure*}

In Figures~\ref{fig:fit-plot_WASP-39b}-\ref{fig:fit-plot_WASP-107b}, we show for each planet one of its data reductions, together with the best-fit model spectra obtained for the baseline retrieval and some selected additional molecules. For clarity, these spectra are smoothed using the scipy gaussian\_filter1d function \citep{Scipy_Virtanen_2020}.

\begin{deluxetable*}{l|cccccccccc}
\renewcommand{\arraystretch}{1.05} 
\setlength{\tabcolsep}{2.2pt}
\tabletypesize{\footnotesize}
\label{tab:summary-BPICS}
\tablecaption{Summary of the 20 molecules reaching $\Delta \overline{BPICS}_{min} \geq 2$ in at least one planet. We present the $\Delta \overline{BPICS}$ values in the same way as the $\ln B$ values in Tables \ref{tab:exploration-uninhabitable} and \ref{tab:summary-molecules}.
This is mostly the same list as in Table~\ref{tab:summary-molecules}, except that C$_8$H$_{16}$ (2,4,4-trimethyl-1-pentene), p-xylene and cyclopentane have been added, whereas chloroethane and diethyl sulfide have been removed.
}
\centering
    \tablehead{
        \colhead{Molecule}  & \colhead{K2-18~b} & \colhead{TOI-270~d} & \colhead{TOI-732~c} & \colhead{WASP-39~b} & \colhead{WASP-17~b}& \colhead{WASP-43~b}&\colhead{GJ~1214~b}&\colhead{L~168-9~b}&\colhead{GJ~367~b}&\colhead{WASP-107~b} }
    \startdata 
        Dimethyl sulfide    & \Bf{2.1}$-$\Bf{6.8} & $-$       & 0.1$-$2.0  &$-$   & $-$     &$-$    & $-$     & $-$ & $-$ & $-$ \\
        Benzonitrile        & $-$         & \Bf{2.8}$-$\Bf{3.4} &$-$      & $-$&$-$    & $<0.2$    & $-$     & $-$ & $-$ & $-$\\
        Cyclohexane         & $<$ 1.8     & $-$       & \Bf{5.7}$-$\Bf{7.3} &$-$ & $-$      &$<$ 0.7    &$-$      & $-$ & $-$ & $-$ \\      
        3-Methylpentane     & -1.2$-$2.4& $-$         & \Bf{4.9}$-$\Bf{5.0} & $-$    & $-$      & $-$       & $-$     & $-$ & $-$ & $-$ \\ 
        2-Butene            & 0.6$-$11.9  & $<0.6$    & \Bf{4.7}$-$\Bf{6.0} & $-$ & $-$      &$-$        & $-$     & $-$ & $-$ & $-$ \\    
        3-Methylhexane      & 0.1$-$6.5   & $-$       & \Bf{3.4}$-$\Bf{4.8} & $-$ & $-$      & $-$       & $-$     & $-$ & $-$ & $-$ \\          
        cis-2-Pentene       & -0.2$-$6.0& $<0.2$      & \Bf{2.6}$-$\Bf{3.0} & $-$    & $-$      & $-$       & $-$     & $-$ & $-$ & $-$ \\ 
        1-Pentene           & $<1.4$      & $<0.6$    & \Bf{2.6}$-$\Bf{2.8} &  $-$ & $-$      & $-$       &$-$  & $-$ & $-$ & $-$ \\      
        Isobutene           & -0.9$-$3.1  & 1.2$-$3.9 & \Bf{2.5}$-$\Bf{5.5} &$-$    &$-$    &$-$        & $-$     & $-$ & $-$ & $-$ \\ 
        Methacrylonitrile   & -1.0$-$3.6  & 1.3$-$2.7 & \Bf{2.4}$-$\Bf{5.0} &$-$&$-$    & $-$    & $-$     & $-$ & $-$ & \Bf{2.8} \\ 
        C$_8$H$_{16}$       & $<0.4$      & $<1.0$    & \Bf{2.3}$-$\Bf{4.6} & $-$    & $-$  & $-$  & $-$ & $-$ & $-$ & $-$ \\   
        Hexane              & 0.1$-$7.4   & $-$       & \Bf{2.2}$-$\Bf{7.2} & $-$    & $-$      &$-$        & $-$     & $-$ & $-$ & $-$ \\    
        H$_2$O              & $-$       & $<0.8^a$  & $<0.1^a$&\Bf{4.4}$-$\Bf{8.6}$^a$ & \Bf{2.9}$-$\Bf{9.4}$^a$ & $-^a$ & $-$& $-$ & $-$ & \Bf{22.9}$^a$ \\ 
        SO$_2$              & $-$       & $-$       & $-$    &\Bf{3.9}$-$\Bf{7.6} &$-$      & $-$      & $-$  & $-$   & $-$ &\Bf{29.5}$^a$ \\
        SO$_2$Cl$_2$        & $-$       & $-$       & $-$    & $-$      &\Bf{2.7}$-$\Bf{3.0}&$-$       & $-$  & $-$   & $-$ & $-$ \\ 
        Cyclopentane   & -1.3$-$3.0& $-$    & 0.2$-$3.3 & $-$      & $-$          &$-$       & \Bf{2.2}  & $-$   & $-$ & $-$ \\ 
        H$_2$S              & $-$       & $-$       & $-$    & $<1.0$   & $-$           & $-$      & $-$  & $-$   & $-$  & \Bf{13.1}$^a$\\
        SO                  & $-$       & $-$       & $-$    & $-$      & $<1.8$        & $-$      & $-$  & $-$   & $-$  & \Bf{9.3}$^a$\\
        PH$_3$              & $<0.8$    & $-$       & $-$    & $-$      & $-$           & $-$      & $-$  & $-$& $-$  & \Bf{8.5}$^a$\\  
        p-Xylene            & $-$       & $-$       & $-$    & $-$      & $-$           & $-$      & $-$  & $-$   & $-$  & \Bf{2.3}\\
    \enddata
    \vspace{2mm}
\end{deluxetable*}

\begin{table*}[]
\centering
\begin{tabular}{ll|ll|ll|ll|ll}
Data  & Case          & \multicolumn{2}{c}{Reference} & \multicolumn{2}{c}{+ Gaussian $M_p$}    &  \multicolumn{2}{c}{$P_{ref} \rightarrow R_p$} &  \multicolumn{2}{c}{Narrow $T_0$}   \\
                  & & $\ln Z$              & $\overline{BPICS}$         & $\ln Z$              & $\overline{BPICS}$ & $\ln Z$              & $\overline{BPICS}$ & $\ln Z$              & $\overline{BPICS}$  \\ \hline
K2-18~b \texttt{JExoRES}    & Baseline  & 213.08 & 208.46 & 213.12 & 207.48 & 211.07 & 208.42 & 212.97 & 208.45 \\
                            & +DMS      & +2.2   & +2.1   & +2.1   & +2.1   & +2.6   & +2.7   & +2.2   & +2.1   \\ \hline
TOI-270~d \texttt{JExoRES}  & Baseline  & 199.31 & 195.84 & 199.32 & 194.78 & 197.93 & 195.61 & 199.25 & 195.93 \\
                            & +Isobutene& +2.3   & +1.2   & +2.3   & +1.2   & +2.2   & +1.4   & +2.1   & +1.2   \\ \hline
WASP-39~b \texttt{Tiberius} & Baseline  & 159.72 & 160.90 & 160.25 & 160.10 & 158.86 & 160.64 & 160.01 & 160.92 \\ 
                            & +SO$_2$   & +6.3   & +7.6   & +5.5   & +7.4   & +5.1   & +7.5   & +6.3   & +7.3   \\ \hline
WASP-107~b \texttt{W24}     & Baseline  & 310.46 & 321.18 & 310.26 & 320.44 & 309.64 & 321.34 & 310.79 & 321.40 \\
                            & -SO$_2$   & -28.4  & -29.5  & -28.2  & -30.1  & -27.6  & -29.7  & -29.0  & -30.4  \\ \hline
\end{tabular}
    \caption{Sensitivity of retrieval results to different prior choices. The reference retrieval setup is the one used throughout this work, described in section~\ref{sec:retrieval_setup} and Appendix~\ref{app:retrieval_priors}. The alternative setups and tested cases are discussed in Appendix~\ref{app:sensitivity-bulk-priors}. For each setup and tested case, we provide the baseline $\ln Z$ and $\overline{BPICS}$ values, as well as the $\ln B$ and $\Delta \overline{BPICS}$ values for adding the listed molecules (or removing SO$_2$ for WASP-107~b).}
\label{tab:sensitivity-bulk}
\end{table*}

\section{List of Species Explored}
\label{app:species-list}
Here we provide the full list of the 160 molecular species used in our study, sorted by opacity source (see section~\ref{subsec:species-explored}). \und{Underlined}: species that were looked for in WASP-107~b (see sections \ref{subsec:exception-WASP-107b} and \ref{subsec:WASP-107b-results}).
Superscript $^b$ indicates species that are included in the baseline retrieval for at least one planet (see section~\ref{subsec:baseline-species} and Appendix~\ref{app:retrieval_priors}).

ExoMol opacities (21):
\und{Chloromethane} (CH$_3$Cl), 
\und{Fluoromethane} (CH$_3$F), 
\und{Methane$^b$} (CH$_4$), 
\und{Carbon monoxide$^b$} (CO), 
\und{Carbon dioxide$^b$} (CO$_2$), 
\und{Carbon monosulfide} (CS), 
\und{Carbon-disulfide} (CS$_2$),
\und{Ethyne} (C$_2$H$_2$), 
\und{Ethylene} (C$_2$H$_4$), 
\und{Hydrogen cyanide} (HCN), 
\und{Thioformaldehyde} (H$_2$CS), 
\und{Water$^b$} (H$_2$O), 
\und{Hydrogen sulfide$^b$} (H$_2$S), 
\und{Ammonia} (NH$_3$), 
\und{Carbonyl sulfide} (OCS), 
\und{Phosphine$^b$} (PH$_3$), 
\und{Silane} (SiH$_4$), 
\und{Silicon monoxide} (SiO), 
\und{Sulfur monoxide$^b$} (SO), 
\und{Sulfur dioxide$^b$} (SO$_2$), 
\und{Sulfur trioxyde} (SO$_3$).

HITRAN line lists: 
\und{Ethane} (C$_2$H$_6$).

HITRAN cross-sections, halogen-bearing (7):
Dichloromethane (CH$_2$Cl$_2$), 
Dichloromethylphosphine (CH$_3$Cl$_2$P),
\und{Bromoethane} (C$_2$H$_5$Br), 
\und{Chloroethane} (C$_2$H$_5$Cl), 
Allyl chloride (C$_3$H$_5$Cl), 
Methacryloyl chloride (C$_4$H$_5$OCl), 
1-Chloropentane (C$_5$H$_{11}$Cl).

HITRAN cross-sections, nitrogen-bearing (13):
Methylamine (CH$_5$N),  
Ethylamine (C$_2$H$_7$N), 
11-Dimethylhydrazine (C$_2$H$_8$N$_2$), 
\und{Ethyl-cyanide} (C$_3$H$_5$N), 
Trimethylamine (C$_3$H$_9$N), 
\und{Methacrylonitrile} (C$_4$H$_5$N), 
Amyl-nitrate (C$_5$H$_{11}$NO$_3$), 
tert-Amylamine (C$_5$H$_{13}$N), 
Pyridine (C$_5$H$_5$N), 
4-Picoline (C$_6$H$_7$N), 
\und{Benzonitrile} (C$_7$H$_5$N), 
o-Toluidine (C$_7$H$_9$N), 
26-Diethylaniline (C$_{10}$H$_{15}$N). 

HITRAN cross-sections, sulfur-bearing (28):
Thiophosegene (CCl$_2$S), 
Perchloromethyl mercaptan (CCl$_4$S), 
Methanesulfonyl-chloride (CH$_3$ClO$_2$S), 
\und{Methanethiol} (CH$_4$S), 
Methyl-isothiocyanate (C$_2$H$_3$NS), 
Ethylene sulfide (C$_2$H$_4$S), 
Dimethyl sulfate (C$_2$H$_6$O$_4$S), 
Dimethyl sulfoxide (C$_2$H$_6$OS), 
Thioglycol (C$_2$H$_6$OS), 
\und{Dimethyl sulfide} (C$_2$H$_6$S), 
\und{Ethyl mercaptan} (C$_2$H$_6$S), 
\und{Dimethyl disulfide} (C$_2$H$_6$S$_2$), 
Propylene sulfide (C$_3$H$_6$S), 
1-Propanethiol (C$_3$H$_8$S), 
2-Propanethiol (C$_3$H$_8$S), 
Diethyl sulfate (C$_4$H$_{10}$O$_4$S), 
2-Methyl-1-propanethiol (C$_4$H$_{10}$S), 
\und{Diethyl sulfide} (C$_4$H$_{10}$S), 
tert-Butylmercaptan (C$_4$H$_{10}$S, 
Thiophene (C$_4$H$_4$S), 
Tetrahydrothiophene (C$_4$H$_8$S), 
Cyclohexanethiol (C$_6$H$_{12}$S), 
Benzenethiol (C$_6$H$_6$S), 
Sulfur hexafluoride (SF$_6$), 
\und{Thionyl fluoride} (SOF$_2$), 
\und{Sulfuryl chloride} (SO$_2$Cl$_2$), 
Sulfuryl fluoride (SO$_2$F$_2$), 
Thiophosphoryl chloride (SPCl$_3$).

HITRAN cross-sections, hydrocarbons (87): 
Allene (C$_3$H$_4$), 
\und{Propyne} (C$_3$H$_4$), 
Cyclopropane (C$_3$H$_6$), 
Propene (C$_3$H$_6$), 
Propane (C$_3$H$_8$), 
\und{Butane} (C$_4$H$_{10}$), 
Isobutane (C$_4$H$_{10}$), 
1-Butyne (C$_4$H$_6$), 
1,3-Butadiene (C$_4$H$_6$), 
1-Butene (C$_4$H$_8$), 
\und{2-Butene} (C$_4$H$_8$), 
\und{Isobutene} (C$_4$H$_8$), 
\und{1-Pentene} (C$_5$H$_{10}$), 
2-Methyl-1-butene (C$_5$H$_{10}$), 
2-Methyl-2-butene (C$_5$H$_{10}$), 
\und{cis-2-Pentene} (C$_5$H$_{10}$), 
trans-2-Pentene (C$_5$H$_{10}$), 
3-Methyl-1-butene (C$_5$H$_{10}$), 
\und{Cyclopentane} (C$_5$H$_{10}$), 
Isopentane (C$_5$H$_{12}$), 
Pentane (C$_5$H$_{12}$), 
Cyclopentene (C$_5$H$_8$), 
\und{Isoprene} (C$_5$H$_8$), 
Cyclohexene (C$_6$H$_{10}$), 
1-Hexene (C$_6$H$_{12}$), 
2-Methyl-1-pentene (C$_6$H$_{12}$), 
2-Methyl-2-pentene (C$_6$H$_{12}$), 
4-Methyl-1-pentene (C$_6$H$_{12}$), 
cis-4-Methyl-2-pentene (C$_6$H$_{12}$), 
\und{Cyclohexane} (C$_6$H$_{12}$), 
22-Dimethylbutane (C$_6$H$_{14}$), 
23-Dimethylbutane (C$_6$H$_{14}$), 
\und{3-Methylpentane} (C$_6$H$_{14}$), 
\und{Hexane} (C$_6$H$_{14}$), 
Isohexane (C$_6$H$_{14}$), 
\und{Benzene} (C$_6$H$_6$), 
Cycloheptene (C$_7$H$_{12}$), 
1-Heptene (C$_7$H$_{14}$), 
Cycloheptane (C$_7$H$_{14}$), 
2,4-Dimethylpentane (C$_7$H$_{16}$), 
\und{3-Methylhexane} (C$_7$H$_{16}$), 
Heptane (C$_7$H$_{16}$), 
Toluene (C$_7$H$_8$), 
Ethyl-benzene (C$_8$H$_{10}$), 
m-Xylene (C$_8$H$_{10}$), 
o-Xylene (C$_8$H$_{10}$), 
\und{p-Xylene} (C$_8$H$_{10}$), 
4-Vinyl-1-cyclohexene (C$_8$H$_{12}$), 
1-Octene (C$_8$H$_{16}$), 
\und{2,4,4-Trimethyl-1-pentene} (C$_8$H$_{16}$), 
2,4,4-Trimethyl-2-pentene (C$_8$H$_{16}$), 
Cyclooctane (C$_8$H$_{16}$), 
Isooctane (C$_8$H$_{18}$), 
Octane (C$_8$H$_{18}$), 
Styrene (C$_8$H$_8$), 
Vinyl toluene (C$_9$H$_{10}$), 
1,2,3-Trimethylbenzene (C$_9$H$_{12}$), 
2-Ethyltoluene (C$_9$H$_{12}$), 
3-Ethyltoluene (C$_9$H$_{12}$), 
4-Ethyltoluene (C$_9$H$_{12}$), 
Cumene (C$_9$H$_{12}$), 
Isocumene (C$_9$H$_{12}$), 
Mesitylene (C$_9$H$_{12}$), 
1-Nonene (C$_9$H$_{18}$),
n-Nonane (C$_9$H$_{20}$), 
Tetralin (C$_{10}$H$_{12}$), 
1,2,3,4-Tetramethylbenzene (C$_{10}$H$_{14}$), 
1,2,3,5-Tetramethylbenzene (C$_{10}$H$_{14}$), 
2-Carene (C$_{10}$H$_{14}$), 
3-Carene (C$_{10}$H$_{14}$), 
sec-Butylbenzene (C$_{10}$H$_{14}$), 
tert-Butylbenzene (C$_{10}$H$_{14}$), 
alpha-Pinene-1S- (C$_{10}$H$_{16}$), 
beta-Pinene-1S- (C$_{10}$H$_{16}$), 
D-Limonene (C$_{10}$H$_{16}$), 
DL-Limonene (C$_{10}$H$_{16}$), 
Myrcene (C$_{10}$H$_{16}$), 
1-Decene (C$_{10}$H$_{20}$), 
Cyclodecane (C$_{10}$H$_{20}$), 
n-Decane (C$_{10}$H$_{22}$), 
Naphthalene (C$_{10}$H$_8$), 
1-Undecene (C$_{11}$H$_{22}$), 
n-Undecane (C$_{11}$H$_{24}$), 
n-Tridecane (C$_{13}$H$_{28}$), 
n-Tetradecane (C$_{14}$H$_{30}$), 
Pentadecane (C$_{15}$H$_{32}$), 
Hexadecane (C$_{16}$H$_{34}$). 

HITRAN cross-sections, others (3):
Dimethyl-carbonate (C$_3$H$_6$O$_3$), 
Isobutenal (C$_4$H$_6$O),
Tetramethylsilane (C$_4$H$_{12}$Si).

\end{document}